\documentclass[longauth]{aa}

\usepackage{graphicx}
\usepackage{natbib}
\usepackage{scalerel}

\usepackage[table]{xcolor}

\usepackage{txfonts}
\usepackage[pdfencoding=auto,psdextra]{hyperref}
\hypersetup{
    colorlinks=true,
    linkcolor=blue,
    filecolor=magenta,      
    urlcolor=blue,
    citecolor=blue
}
\makeatletter
\renewcommand*\aa@pageof{, page \thepage{} of \pageref*{LastPage}}
\makeatother

\usepackage[utf8]{inputenc}

\usepackage[switch, modulo]{lineno}
              
\renewcommand{\linenumbers}[0]{}

\usepackage{euclid}
\usepackage{multicol}

\newcommand*{\Ecker}{Euclid Collaboration: Ecker et al. (in prep.)}
\newcommand{\ignoreme}[1]{}

\usepackage{xcolor}
\definecolor{nnone}{RGB}{31,119,180}     
\definecolor{nntwo}{RGB}{255,127,14}     
\definecolor{nnthree}{RGB}{44,160,44}    
\definecolor{nnfour}{RGB}{214,39,40}     
\definecolor{nnfive}{RGB}{148,103,189}   

\definecolor{gradeA}{RGB}{0,100,0}       
\definecolor{gradeB}{RGB}{34,139,34}     
\definecolor{gradeC}{RGB}{255,140,0}     
\definecolor{gradeX}{RGB}{214,39,40}     

\begin{document}
%
%

\title{Euclid Data Release 1 (DR1)}
\subtitle{Learning machine learning: Investigating the complementarity and biases of strong gravitational lens finding networks on \Euclid data}      



\newcommand{\orcid}[1]{} 
\author{Euclid Collaboration: S.~H.~Vincken\orcid{0009-0005-7305-2359}\thanks{\email{saamie.vincken@fhnw.ch}}\inst{\ref{aff1},\ref{aff2},\ref{aff3}}
\and N.~E.~P.~Lines\orcid{0009-0004-7751-1914}\inst{\ref{aff1}}
\and S.~Schuldt\orcid{0000-0003-2497-6334}\inst{\ref{aff4},\ref{aff5},\ref{aff6}}
\and P.~Holloway\orcid{0009-0002-8896-6100}\inst{\ref{aff1}}
\and T.~E.~Collett\orcid{0000-0001-5564-3140}\inst{\ref{aff1}}
\and M.~Fogliardi\orcid{0009-0006-4964-5311}\inst{\ref{aff7}}
\and A.~Manj\'on-Garc\'ia\orcid{0000-0002-7413-8825}\inst{\ref{aff8}}
\and J.~Wilde\orcid{0000-0002-4460-7379}\inst{\ref{aff9}}
\and A.~Verma\orcid{0000-0002-0730-0781}\inst{\ref{aff10}}
\and K.~Rojas\orcid{0000-0003-1391-6854}\inst{\ref{aff11}}
\and R.~B.~Metcalf\orcid{0000-0003-3167-2574}\inst{\ref{aff12},\ref{aff13}}
\and N.~Wielders\orcid{0009-0001-8110-819X}\inst{\ref{aff14}}
\and J.~W.~Nightingale\orcid{0000-0002-8987-7401}\inst{\ref{aff14}}
\and A.~Nersesian\orcid{0000-0001-6843-409X}\inst{\ref{aff15},\ref{aff16}}
\and F.~Courbin\orcid{0000-0003-0758-6510}\inst{\ref{aff9},\ref{aff17},\ref{aff18}}
\and S.~Serjeant\orcid{0000-0002-0517-7943}\inst{\ref{aff19}}
\and L.~A.~Moustakas\orcid{0000-0003-3030-2360}\inst{\ref{aff20}}
\and D.~Stern\orcid{0000-0003-2686-9241}\inst{\ref{aff20}}
\and S.~Kruk\orcid{0000-0001-8010-8879}\inst{\ref{aff21}}
\and D.~Sluse\orcid{0000-0001-6116-2095}\inst{\ref{aff15}}
\and G.~Despali\orcid{0000-0001-6150-4112}\inst{\ref{aff12},\ref{aff13},\ref{aff22}}
\and R.~Gavazzi\orcid{0000-0002-5540-6935}\inst{\ref{aff23},\ref{aff24}}
\and H.~Degaudenzi\orcid{0000-0002-5887-6799}\inst{\ref{aff25}}
\and M.~Grespan\orcid{0000-0002-8213-0217}\inst{\ref{aff10},\ref{aff26}}
\and G.~Granata\orcid{0000-0002-9512-3788}\inst{\ref{aff1}}
\and D.~P.~Johnson\orcid{0000-0002-0311-2513}\inst{\ref{aff27}}
\and C.~M.~O'Riordan\orcid{0000-0003-2227-1998}\inst{\ref{aff9}}
\and M.~Millon\orcid{0000-0001-7051-497X}\inst{\ref{aff28}}
\and D.~J.~Bowden\orcid{0009-0008-6114-1401}\inst{\ref{aff10}}
\and J.~Pearson\orcid{0000-0001-8555-8561}\inst{\ref{aff19}}
\and N.~Massari\orcid{0009-0006-8421-930X}\inst{\ref{aff23}}
\and N.~Aghanim\orcid{0000-0002-6688-8992}\inst{\ref{aff29}}
\and S.~Andreon\orcid{0000-0002-2041-8784}\inst{\ref{aff30}}
\and N.~Auricchio\orcid{0000-0003-4444-8651}\inst{\ref{aff13}}
\and C.~Baccigalupi\orcid{0000-0002-8211-1630}\inst{\ref{aff31},\ref{aff32},\ref{aff33},\ref{aff34}}
\and M.~Baldi\orcid{0000-0003-4145-1943}\inst{\ref{aff35},\ref{aff13},\ref{aff22}}
\and A.~Balestra\orcid{0000-0002-6967-261X}\inst{\ref{aff36}}
\and S.~Bardelli\orcid{0000-0002-8900-0298}\inst{\ref{aff13}}
\and P.~Battaglia\orcid{0000-0002-7337-5909}\inst{\ref{aff13}}
\and A.~Biviano\orcid{0000-0002-0857-0732}\inst{\ref{aff32},\ref{aff31}}
\and M.~Bolzonella\orcid{0000-0003-3278-4607}\inst{\ref{aff13}}
\and E.~Branchini\orcid{0000-0002-0808-6908}\inst{\ref{aff37},\ref{aff38},\ref{aff30}}
\and M.~Brescia\orcid{0000-0001-9506-5680}\inst{\ref{aff39},\ref{aff40}}
\and S.~Camera\orcid{0000-0003-3399-3574}\inst{\ref{aff41},\ref{aff42},\ref{aff43}}
\and G.~Ca\~nas-Herrera\orcid{0000-0003-2796-2149}\inst{\ref{aff44}}
\and V.~Capobianco\orcid{0000-0002-3309-7692}\inst{\ref{aff43}}
\and C.~Carbone\orcid{0000-0003-0125-3563}\inst{\ref{aff6}}
\and J.~Carretero\orcid{0000-0002-3130-0204}\inst{\ref{aff45},\ref{aff46}}
\and F.~J.~Castander\orcid{0000-0001-7316-4573}\inst{\ref{aff47},\ref{aff48}}
\and M.~Castellano\orcid{0000-0001-9875-8263}\inst{\ref{aff49}}
\and G.~Castignani\orcid{0000-0001-6831-0687}\inst{\ref{aff13}}
\and S.~Cavuoti\orcid{0000-0002-3787-4196}\inst{\ref{aff40},\ref{aff50}}
\and A.~Cimatti\inst{\ref{aff51}}
\and C.~Colodro-Conde\inst{\ref{aff52}}
\and G.~Congedo\orcid{0000-0003-2508-0046}\inst{\ref{aff53}}
\and L.~Conversi\orcid{0000-0002-6710-8476}\inst{\ref{aff54},\ref{aff21}}
\and Y.~Copin\orcid{0000-0002-5317-7518}\inst{\ref{aff55}}
\and H.~M.~Courtois\orcid{0000-0003-0509-1776}\inst{\ref{aff56}}
\and M.~Cropper\orcid{0000-0003-4571-9468}\inst{\ref{aff57}}
\and A.~Da~Silva\orcid{0000-0002-6385-1609}\inst{\ref{aff58},\ref{aff59}}
\and G.~De~Lucia\orcid{0000-0002-6220-9104}\inst{\ref{aff32}}
\and C.~Dolding\orcid{0009-0003-7199-6108}\inst{\ref{aff57}}
\and H.~Dole\orcid{0000-0002-9767-3839}\inst{\ref{aff29}}
\and F.~Dubath\orcid{0000-0002-6533-2810}\inst{\ref{aff25}}
\and X.~Dupac\inst{\ref{aff21}}
\and S.~Dusini\orcid{0000-0002-1128-0664}\inst{\ref{aff60}}
\and S.~Escoffier\orcid{0000-0002-2847-7498}\inst{\ref{aff61}}
\and M.~Farina\orcid{0000-0002-3089-7846}\inst{\ref{aff62}}
\and R.~Farinelli\inst{\ref{aff13}}
\and F.~Faustini\orcid{0000-0001-6274-5145}\inst{\ref{aff49}}
\and S.~Ferriol\inst{\ref{aff55}}
\and S.~Fotopoulou\orcid{0000-0002-9686-254X}\inst{\ref{aff63}}
\and M.~Frailis\orcid{0000-0002-7400-2135}\inst{\ref{aff32}}
\and E.~Franceschi\orcid{0000-0002-0585-6591}\inst{\ref{aff13}}
\and M.~Fumana\orcid{0000-0001-6787-5950}\inst{\ref{aff6}}
\and L.~Gabarra\orcid{0000-0002-8486-8856}\inst{\ref{aff10}}
\and S.~Galeotta\orcid{0000-0002-3748-5115}\inst{\ref{aff32}}
\and K.~George\orcid{0000-0002-1734-8455}\inst{\ref{aff64}}
\and B.~Gillis\orcid{0000-0002-4478-1270}\inst{\ref{aff53}}
\and C.~Giocoli\orcid{0000-0002-9590-7961}\inst{\ref{aff13},\ref{aff22}}
\and J.~Gracia-Carpio\orcid{0000-0003-4689-3134}\inst{\ref{aff65}}
\and A.~Grazian\orcid{0000-0002-5688-0663}\inst{\ref{aff36}}
\and F.~Grupp\inst{\ref{aff65},\ref{aff66}}
\and W.~G.~Hartley\orcid{0000-0002-4563-5101}\inst{\ref{aff25}}
\and S.~V.~H.~Haugan\orcid{0000-0001-9648-7260}\inst{\ref{aff67}}
\and S.~Hemmati\orcid{0000-0003-2226-5395}\inst{\ref{aff68}}
\and W.~Holmes\orcid{0009-0007-8554-4646}\inst{\ref{aff20}}
\and I.~M.~Hook\orcid{0000-0002-2960-978X}\inst{\ref{aff69}}
\and F.~Hormuth\inst{\ref{aff70}}
\and A.~Hornstrup\orcid{0000-0002-3363-0936}\inst{\ref{aff71},\ref{aff72}}
\and K.~Jahnke\orcid{0000-0003-3804-2137}\inst{\ref{aff73}}
\and M.~Jhabvala\inst{\ref{aff74}}
\and B.~Joachimi\orcid{0000-0001-7494-1303}\inst{\ref{aff75}}
\and S.~Kermiche\orcid{0000-0002-0302-5735}\inst{\ref{aff61}}
\and A.~Kiessling\orcid{0000-0002-2590-1273}\inst{\ref{aff20}}
\and B.~Kubik\orcid{0009-0006-5823-4880}\inst{\ref{aff55}}
\and M.~K\"ummel\orcid{0000-0003-2791-2117}\inst{\ref{aff66}}
\and M.~Kunz\orcid{0000-0002-3052-7394}\inst{\ref{aff28}}
\and H.~Kurki-Suonio\orcid{0000-0002-4618-3063}\inst{\ref{aff5},\ref{aff76}}
\and A.~M.~C.~Le~Brun\orcid{0000-0002-0936-4594}\inst{\ref{aff77}}
\and S.~Ligori\orcid{0000-0003-4172-4606}\inst{\ref{aff43}}
\and P.~B.~Lilje\orcid{0000-0003-4324-7794}\inst{\ref{aff67}}
\and V.~Lindholm\orcid{0000-0003-2317-5471}\inst{\ref{aff5},\ref{aff76}}
\and I.~Lloro\orcid{0000-0001-5966-1434}\inst{\ref{aff78}}
\and M.~Magliocchetti\orcid{0000-0001-9158-4838}\inst{\ref{aff62}}
\and G.~Mainetti\orcid{0000-0003-2384-2377}\inst{\ref{aff79}}
\and O.~Mansutti\orcid{0000-0001-5758-4658}\inst{\ref{aff32}}
\and O.~Marggraf\orcid{0000-0001-7242-3852}\inst{\ref{aff80}}
\and M.~Martinelli\orcid{0000-0002-6943-7732}\inst{\ref{aff49},\ref{aff81}}
\and N.~Martinet\orcid{0000-0003-2786-7790}\inst{\ref{aff23}}
\and F.~Marulli\orcid{0000-0002-8850-0303}\inst{\ref{aff12},\ref{aff13},\ref{aff22}}
\and R.~J.~Massey\orcid{0000-0002-6085-3780}\inst{\ref{aff82}}
\and N.~Mauri\orcid{0000-0001-8196-1548}\inst{\ref{aff51},\ref{aff22}}
\and E.~Medinaceli\orcid{0000-0002-4040-7783}\inst{\ref{aff13}}
\and S.~Mei\orcid{0000-0002-2849-559X}\inst{\ref{aff83},\ref{aff84}}
\and M.~Melchior\inst{\ref{aff3}}
\and M.~Meneghetti\orcid{0000-0003-1225-7084}\inst{\ref{aff13},\ref{aff22}}
\and E.~Merlin\orcid{0000-0001-6870-8900}\inst{\ref{aff36}}
\and G.~Meylan\orcid{0000-0001-6503-0209}\inst{\ref{aff85}}
\and P.~Monaco\orcid{0000-0003-2083-7564}\inst{\ref{aff86},\ref{aff32},\ref{aff33},\ref{aff31}}
\and A.~Mora\orcid{0000-0002-1922-8529}\inst{\ref{aff87}}
\and M.~Moresco\orcid{0000-0002-7616-7136}\inst{\ref{aff12},\ref{aff13}}
\and C.~Moretti\orcid{0000-0003-3314-8936}\inst{\ref{aff32},\ref{aff31},\ref{aff33}}
\and L.~Moscardini\orcid{0000-0002-3473-6716}\inst{\ref{aff12},\ref{aff13},\ref{aff22}}
\and C.~Neissner\orcid{0000-0001-8524-4968}\inst{\ref{aff88},\ref{aff46}}
\and R.~C.~Nichol\orcid{0000-0003-0939-6518}\inst{\ref{aff89}}
\and S.-M.~Niemi\orcid{0009-0005-0247-0086}\inst{\ref{aff90}}
\and C.~Padilla\orcid{0000-0001-7951-0166}\inst{\ref{aff88}}
\and S.~Paltani\orcid{0000-0002-8108-9179}\inst{\ref{aff25}}
\and F.~Pasian\orcid{0000-0002-4869-3227}\inst{\ref{aff32}}
\and W.~J.~Percival\orcid{0000-0002-0644-5727}\inst{\ref{aff91},\ref{aff92},\ref{aff93}}
\and V.~Pettorino\orcid{0000-0002-4203-9320}\inst{\ref{aff90}}
\and A.~Pezzotta\orcid{0000-0003-0726-2268}\inst{\ref{aff30}}
\and S.~Pires\orcid{0000-0002-0249-2104}\inst{\ref{aff94}}
\and G.~Polenta\orcid{0000-0003-4067-9196}\inst{\ref{aff95}}
\and M.~Poncet\inst{\ref{aff96}}
\and L.~A.~Popa\inst{\ref{aff97}}
\and F.~Raison\orcid{0000-0002-7819-6918}\inst{\ref{aff65}}
\and R.~Rebolo\orcid{0000-0003-3767-7085}\inst{\ref{aff52},\ref{aff98},\ref{aff99}}
\and A.~Renzi\orcid{0000-0001-9856-1970}\inst{\ref{aff100},\ref{aff60},\ref{aff13}}
\and J.~Rhodes\orcid{0000-0002-4485-8549}\inst{\ref{aff20}}
\and G.~Riccio\inst{\ref{aff40}}
\and I.~Risso\orcid{0000-0003-2525-7761}\inst{\ref{aff37},\ref{aff38},\ref{aff30}}
\and F.~Rizzo\orcid{0000-0002-9407-585X}\inst{\ref{aff32}}
\and E.~Romelli\orcid{0000-0003-3069-9222}\inst{\ref{aff32}}
\and M.~Roncarelli\orcid{0000-0001-9587-7822}\inst{\ref{aff13}}
\and B.~Rusholme\orcid{0000-0001-7648-4142}\inst{\ref{aff68}}
\and R.~Saglia\orcid{0000-0003-0378-7032}\inst{\ref{aff66},\ref{aff65}}
\and Z.~Sakr\orcid{0000-0002-4823-3757}\inst{\ref{aff101},\ref{aff102},\ref{aff103}}
\and A.~G.~S\'anchez\orcid{0000-0003-1198-831X}\inst{\ref{aff65}}
\and D.~Sapone\orcid{0000-0001-7089-4503}\inst{\ref{aff104}}
\and B.~Sartoris\orcid{0000-0003-1337-5269}\inst{\ref{aff65},\ref{aff32}}
\and P.~Schneider\orcid{0000-0001-8561-2679}\inst{\ref{aff80}}
\and T.~Schrabback\orcid{0000-0002-6987-7834}\inst{\ref{aff105}}
\and A.~Secroun\orcid{0000-0003-0505-3710}\inst{\ref{aff61}}
\and E.~Sihvola\orcid{0000-0003-1804-7715}\inst{\ref{aff106}}
\and C.~Sirignano\orcid{0000-0002-0995-7146}\inst{\ref{aff100},\ref{aff60}}
\and G.~Sirri\orcid{0000-0003-2626-2853}\inst{\ref{aff22}}
\and L.~Stanco\orcid{0000-0002-9706-5104}\inst{\ref{aff60}}
\and J.-L.~Starck\orcid{0000-0003-2177-7794}\inst{\ref{aff94}}
\and P.~Tallada-Cresp\'{i}\orcid{0000-0002-1336-8328}\inst{\ref{aff45},\ref{aff46}}
\and A.~N.~Taylor\inst{\ref{aff53}}
\and I.~Tereno\orcid{0000-0002-4537-6218}\inst{\ref{aff58},\ref{aff107}}
\and S.~Toft\orcid{0000-0003-3631-7176}\inst{\ref{aff108},\ref{aff109}}
\and R.~Toledo-Moreo\orcid{0000-0002-2997-4859}\inst{\ref{aff110},\ref{aff111}}
\and F.~Torradeflot\orcid{0000-0003-1160-1517}\inst{\ref{aff46},\ref{aff45}}
\and I.~Tutusaus\orcid{0000-0002-3199-0399}\inst{\ref{aff47},\ref{aff48},\ref{aff102}}
\and J.~Valiviita\orcid{0000-0001-6225-3693}\inst{\ref{aff5},\ref{aff76}}
\and T.~Vassallo\orcid{0000-0001-6512-6358}\inst{\ref{aff32},\ref{aff64}}
\and G.~Verdoes~Kleijn\orcid{0000-0001-5803-2580}\inst{\ref{aff112}}
\and A.~Veropalumbo\orcid{0000-0003-2387-1194}\inst{\ref{aff30},\ref{aff38},\ref{aff37}}
\and Y.~Wang\orcid{0000-0002-4749-2984}\inst{\ref{aff68}}
\and J.~Weller\orcid{0000-0002-8282-2010}\inst{\ref{aff66},\ref{aff65}}
\and A.~Zacchei\orcid{0000-0003-0396-1192}\inst{\ref{aff32},\ref{aff31}}
\and G.~Zamorani\orcid{0000-0002-2318-301X}\inst{\ref{aff13}}
\and E.~Zucca\orcid{0000-0002-5845-8132}\inst{\ref{aff13}}
\and V.~Allevato\orcid{0000-0001-7232-5152}\inst{\ref{aff40}}
\and M.~Ballardini\orcid{0000-0003-4481-3559}\inst{\ref{aff7},\ref{aff113},\ref{aff13}}
\and P.~Bergamini\orcid{0000-0003-1383-9414}\inst{\ref{aff13}}
\and S.~Borgani\orcid{0000-0001-6151-6439}\inst{\ref{aff86},\ref{aff31},\ref{aff32},\ref{aff33},\ref{aff114}}
\and E.~Bozzo\orcid{0000-0002-8201-1525}\inst{\ref{aff25}}
\and C.~Burigana\orcid{0000-0002-3005-5796}\inst{\ref{aff115},\ref{aff116}}
\and R.~Cabanac\orcid{0000-0001-6679-2600}\inst{\ref{aff102}}
\and M.~Calabrese\orcid{0000-0002-2637-2422}\inst{\ref{aff117},\ref{aff6}}
\and A.~Cappi\inst{\ref{aff118},\ref{aff13}}
\and F.~Caro\orcid{0009-0003-1053-0507}\inst{\ref{aff49}}
\and T.~Castro\orcid{0000-0002-6292-3228}\inst{\ref{aff119},\ref{aff32},\ref{aff31}}
\and J.~A.~Escartin~Vigo\inst{\ref{aff65}}
\and J.~Garc\'ia-Bellido\orcid{0000-0002-9370-8360}\inst{\ref{aff101}}
\and T.~Gasparetto\orcid{0000-0002-7913-4866}\inst{\ref{aff49}}
\and V.~Gautard\inst{\ref{aff120}}
\and A.~Loureiro\orcid{0000-0002-4371-0876}\inst{\ref{aff121},\ref{aff122}}
\and R.~Maoli\orcid{0000-0002-6065-3025}\inst{\ref{aff123},\ref{aff49}}
\and A.~A.~Nucita\inst{\ref{aff124},\ref{aff125},\ref{aff126}}
\and M.~P\"ontinen\orcid{0000-0001-5442-2530}\inst{\ref{aff5}}
\and E.~Sarpa\orcid{0000-0002-1256-655X}\inst{\ref{aff32}}
\and V.~Scottez\orcid{0009-0008-3864-940X}\inst{\ref{aff127},\ref{aff128}}
\and M.~Sereno\orcid{0000-0003-0302-0325}\inst{\ref{aff13},\ref{aff22}}
\and M.~Tenti\orcid{0000-0002-4254-5901}\inst{\ref{aff22}}
\and M.~Tucci\inst{\ref{aff25}}
\and M.~Viel\orcid{0000-0002-2642-5707}\inst{\ref{aff31},\ref{aff32},\ref{aff34},\ref{aff33},\ref{aff114}}
\and M.~Wiesmann\orcid{0009-0000-8199-5860}\inst{\ref{aff67}}
\and J.~A.~Acevedo~Barroso\orcid{0000-0002-9654-1711}\inst{\ref{aff20}}
\and Y.~Akrami\orcid{0000-0002-2407-7956}\inst{\ref{aff101},\ref{aff129}}
\and I.~T.~Andika\orcid{0000-0001-6102-9526}\inst{\ref{aff66}}
\and S.~Anselmi\orcid{0000-0002-3579-9583}\inst{\ref{aff60},\ref{aff100},\ref{aff130}}
\and M.~Archidiacono\orcid{0000-0003-4952-9012}\inst{\ref{aff131},\ref{aff132}}
\and G.~Aric\`o\orcid{0000-0002-2802-2928}\inst{\ref{aff22}}
\and F.~Atrio-Barandela\orcid{0000-0002-2130-2513}\inst{\ref{aff133}}
\and E.~Aubourg\orcid{0000-0002-5592-023X}\inst{\ref{aff83},\ref{aff134}}
\and M.~Baes\orcid{0000-0002-3930-2757}\inst{\ref{aff16}}
\and L.~Bazzanini\orcid{0000-0003-0727-0137}\inst{\ref{aff7},\ref{aff13}}
\and D.~Bertacca\orcid{0000-0002-2490-7139}\inst{\ref{aff100},\ref{aff36},\ref{aff60}}
\and M.~Bethermin\orcid{0000-0002-3915-2015}\inst{\ref{aff135}}
\and F.~Beutler\orcid{0000-0003-0467-5438}\inst{\ref{aff53}}
\and A.~Blanchard\orcid{0000-0001-8555-9003}\inst{\ref{aff102}}
\and L.~Blot\orcid{0000-0002-9622-7167}\inst{\ref{aff136},\ref{aff77}}
\and M.~L.~Brown\orcid{0000-0002-0370-8077}\inst{\ref{aff137}}
\and S.~Bruton\orcid{0000-0002-6503-5218}\inst{\ref{aff138}}
\and A.~Calabro\orcid{0000-0003-2536-1614}\inst{\ref{aff49}}
\and B.~Camacho~Quevedo\orcid{0000-0002-8789-4232}\inst{\ref{aff31},\ref{aff34},\ref{aff32}}
\and C.~S.~Carvalho\inst{\ref{aff107}}
\and Y.~Charles\orcid{0009-0000-3636-834X}\inst{\ref{aff23}}
\and F.~Cogato\orcid{0000-0003-4632-6113}\inst{\ref{aff12},\ref{aff13}}
\and S.~Conseil\orcid{0000-0002-3657-4191}\inst{\ref{aff55}}
\and A.~R.~Cooray\orcid{0000-0002-3892-0190}\inst{\ref{aff139}}
\and P.~Corcho-Caballero\orcid{0000-0001-6327-7080}\inst{\ref{aff112}}
\and B.~Csizi\orcid{0000-0003-3227-6581}\inst{\ref{aff105}}
\and O.~Cucciati\orcid{0000-0002-9336-7551}\inst{\ref{aff13}}
\and S.~Davini\orcid{0000-0003-3269-1718}\inst{\ref{aff38}}
\and T.~de~Boer\orcid{0000-0001-5486-2747}\inst{\ref{aff140}}
\and G.~Desprez\orcid{0000-0001-8325-1742}\inst{\ref{aff112}}
\and A.~D\'iaz-S\'anchez\orcid{0000-0003-0748-4768}\inst{\ref{aff8}}
\and S.~Di~Domizio\orcid{0000-0003-2863-5895}\inst{\ref{aff37},\ref{aff38}}
\and J.~M.~Diego\orcid{0000-0001-9065-3926}\inst{\ref{aff141}}
\and P.-A.~Duc\orcid{0000-0003-3343-6284}\inst{\ref{aff135}}
\and V.~Duret\orcid{0009-0009-0383-4960}\inst{\ref{aff61}}
\and M.~Y.~Elkhashab\orcid{0000-0001-9306-2603}\inst{\ref{aff32},\ref{aff33},\ref{aff86},\ref{aff31}}
\and Y.~Fang\orcid{0000-0002-0334-6950}\inst{\ref{aff66}}
\and A.~Farina\orcid{0009-0000-3420-929X}\inst{\ref{aff30},\ref{aff38}}
\and A.~G.~Ferrari\orcid{0009-0005-5266-4110}\inst{\ref{aff22}}
\and A.~Finoguenov\orcid{0000-0002-4606-5403}\inst{\ref{aff5}}
\and A.~Franco\orcid{0000-0002-4761-366X}\inst{\ref{aff124},\ref{aff125},\ref{aff126}}
\and Y.~Fu\orcid{0000-0002-0759-0504}\inst{\ref{aff44},\ref{aff112}}
\and K.~Ganga\orcid{0000-0001-8159-8208}\inst{\ref{aff83}}
\and E.~Gaztanaga\orcid{0000-0001-9632-0815}\inst{\ref{aff47},\ref{aff48},\ref{aff1}}
\and Z.~Ghaffari\orcid{0000-0002-6467-8078}\inst{\ref{aff32},\ref{aff31}}
\and F.~Giacomini\orcid{0000-0002-3129-2814}\inst{\ref{aff22}}
\and F.~Gianotti\orcid{0000-0003-4666-119X}\inst{\ref{aff13}}
\and G.~Gozaliasl\orcid{0000-0002-0236-919X}\inst{\ref{aff142},\ref{aff5}}
\and A.~Gruppuso\orcid{0000-0001-9272-5292}\inst{\ref{aff13},\ref{aff22}}
\and M.~Guidi\orcid{0000-0001-9408-1101}\inst{\ref{aff35},\ref{aff13}}
\and C.~M.~Gutierrez\orcid{0000-0001-7854-783X}\inst{\ref{aff52},\ref{aff99}}
\and A.~Hall\orcid{0000-0002-3139-8651}\inst{\ref{aff53}}
\and C.~Hern\'andez-Monteagudo\orcid{0000-0001-5471-9166}\inst{\ref{aff99},\ref{aff52}}
\and H.~Hildebrandt\orcid{0000-0002-9814-3338}\inst{\ref{aff143}}
\and J.~Hjorth\orcid{0000-0002-4571-2306}\inst{\ref{aff144}}
\and M.~Jauzac\orcid{0000-0003-1974-8732}\inst{\ref{aff145},\ref{aff82},\ref{aff146},\ref{aff147},\ref{aff102}}
\and J.~J.~E.~Kajava\orcid{0000-0002-3010-8333}\inst{\ref{aff148},\ref{aff4},\ref{aff149}}
\and Y.~Kang\orcid{0009-0000-8588-7250}\inst{\ref{aff105}}
\and V.~Kansal\orcid{0000-0002-4008-6078}\inst{\ref{aff150},\ref{aff151}}
\and D.~Karagiannis\orcid{0000-0002-4927-0816}\inst{\ref{aff7},\ref{aff152}}
\and J.~Kim\orcid{0000-0003-2776-2761}\inst{\ref{aff10}}
\and C.~C.~Kirkpatrick\inst{\ref{aff106}}
\and A.~Kov\'acs\orcid{0000-0002-5825-579X}\inst{\ref{aff153},\ref{aff154}}
\and I.~Kova{\v{c}}i{\'{c}}\orcid{0000-0001-6751-3263}\inst{\ref{aff16}}
\and K.~Koyama\orcid{0000-0001-6727-6915}\inst{\ref{aff1}}
\and M.~C.~Lam\orcid{0000-0002-9347-2298}\inst{\ref{aff53}}
\and M.~Lattanzi\orcid{0000-0003-1059-2532}\inst{\ref{aff113}}
\and L.~Legrand\orcid{0000-0003-0610-5252}\inst{\ref{aff155},\ref{aff29}}
\and G.~F.~Lesci\orcid{0000-0002-4607-2830}\inst{\ref{aff12},\ref{aff13}}
\and J.~Lesgourgues\orcid{0000-0001-7627-353X}\inst{\ref{aff156}}
\and L.~Linke\orcid{0000-0002-2622-8113}\inst{\ref{aff105}}
\and G.~Maggio\orcid{0000-0003-4020-4836}\inst{\ref{aff32}}
\and F.~Mannucci\orcid{0000-0002-4803-2381}\inst{\ref{aff157}}
\and F.~R.~Marleau\orcid{0000-0002-1442-2947}\inst{\ref{aff105}}
\and C.~J.~A.~P.~Martins\orcid{0000-0002-4886-9261}\inst{\ref{aff158},\ref{aff159}}
\and M.~Miluzio\inst{\ref{aff21},\ref{aff160}}
\and G.~Morgante\inst{\ref{aff13}}
\and S.~Nadathur\orcid{0000-0001-9070-3102}\inst{\ref{aff1}}
\and K.~Naidoo\orcid{0000-0002-9182-1802}\inst{\ref{aff1},\ref{aff73}}
\and P.~Natoli\orcid{0000-0003-0126-9100}\inst{\ref{aff7},\ref{aff113}}
\and A.~Navarro-Alsina\orcid{0000-0002-3173-2592}\inst{\ref{aff80}}
\and S.~Nesseris\orcid{0000-0002-0567-0324}\inst{\ref{aff101}}
\and L.~Nicastro\orcid{0000-0001-8534-6788}\inst{\ref{aff13}}
\and M.~Oguri\orcid{0000-0003-3484-399X}\inst{\ref{aff161},\ref{aff162}}
\and F.~Oppizzi\orcid{0000-0003-3904-8370}\inst{\ref{aff38},\ref{aff60},\ref{aff100}}
\and F.~Pace\orcid{0000-0001-8039-0480}\inst{\ref{aff41},\ref{aff42},\ref{aff43},\ref{aff59}}
\and D.~Paoletti\orcid{0000-0003-4761-6147}\inst{\ref{aff13},\ref{aff116}}
\and F.~Passalacqua\orcid{0000-0002-8606-4093}\inst{\ref{aff60}}
\and K.~Paterson\orcid{0000-0001-8340-3486}\inst{\ref{aff73}}
\and L.~Patrizii\inst{\ref{aff22}}
\and C.~Pattison\orcid{0000-0003-3272-2617}\inst{\ref{aff1}}
\and R.~Paviot\orcid{0009-0002-8108-3460}\inst{\ref{aff94},\ref{aff96}}
\and A.~Pisani\orcid{0000-0002-6146-4437}\inst{\ref{aff61}}
\and D.~Potter\orcid{0000-0002-0757-5195}\inst{\ref{aff163}}
\and G.~W.~Pratt\inst{\ref{aff94}}
\and S.~Quai\orcid{0000-0002-0449-8163}\inst{\ref{aff12},\ref{aff13}}
\and M.~Radovich\orcid{0000-0002-3585-866X}\inst{\ref{aff36}}
\and G.~Rodighiero\orcid{0000-0002-9415-2296}\inst{\ref{aff100},\ref{aff36}}
\and W.~Roster\orcid{0000-0002-9149-6528}\inst{\ref{aff65}}
\and S.~Sacquegna\orcid{0000-0002-8433-6630}\inst{\ref{aff164}}
\and M.~Sahl\'en\orcid{0000-0003-0973-4804}\inst{\ref{aff165}}
\and D.~B.~Sanders\orcid{0000-0002-1233-9998}\inst{\ref{aff140}}
\and A.~Schneider\orcid{0000-0001-7055-8104}\inst{\ref{aff163}}
\and D.~Sciotti\orcid{0009-0008-4519-2620}\inst{\ref{aff49},\ref{aff81}}
\and E.~Sellentin\orcid{0009-0002-2655-3458}\inst{\ref{aff166},\ref{aff44}}
\and L.~C.~Smith\orcid{0000-0002-3259-2771}\inst{\ref{aff167}}
\and J.~G.~Sorce\orcid{0000-0002-2307-2432}\inst{\ref{aff168},\ref{aff29}}
\and I.~Szapudi\orcid{0000-0003-2274-0301}\inst{\ref{aff140}}
\and M.~Talia\orcid{0000-0003-4352-2063}\inst{\ref{aff12},\ref{aff13}}
\and K.~Tanidis\orcid{0000-0001-9843-5130}\inst{\ref{aff169}}
\and C.~Tao\orcid{0000-0001-7961-8177}\inst{\ref{aff61}}
\and G.~Testera\orcid{0000-0003-2970-766X}\inst{\ref{aff38}}
\and R.~Teyssier\orcid{0000-0001-7689-0933}\inst{\ref{aff170}}
\and S.~Tosi\orcid{0000-0002-7275-9193}\inst{\ref{aff37},\ref{aff30},\ref{aff38}}
\and A.~Troja\orcid{0000-0003-0239-4595}\inst{\ref{aff32}}
\and C.~Uhlemann\orcid{0000-0001-7831-1579}\inst{\ref{aff171},\ref{aff14}}
\and C.~Valieri\inst{\ref{aff22}}
\and A.~Venhola\orcid{0000-0001-6071-4564}\inst{\ref{aff172}}
\and D.~Vergani\orcid{0000-0003-0898-2216}\inst{\ref{aff13}}
\and G.~Vernardos\orcid{0000-0001-8554-7248}\inst{\ref{aff173},\ref{aff174}}
\and G.~Verza\orcid{0000-0002-1886-8348}\thanks{Deceased}\inst{\ref{aff175},\ref{aff176}}
\and E.~Vilenius\orcid{0000-0002-6184-7681}\inst{\ref{aff57}}
\and S.~Vinciguerra\orcid{0009-0005-4018-3184}\inst{\ref{aff23}}
\and M.~von~Wietersheim-Kramsta\orcid{0000-0003-4986-5091}\inst{\ref{aff82},\ref{aff145}}
\and N.~A.~Walton\orcid{0000-0003-3983-8778}\inst{\ref{aff167}}
\and L.~Wang\orcid{0000-0002-6736-9158}\inst{\ref{aff177},\ref{aff112}}
\and A.~H.~Wright\orcid{0000-0001-7363-7932}\inst{\ref{aff143}}
\and H.~W.~Yeung\orcid{0000-0002-4993-9014}\inst{\ref{aff53}}
\and D.~Scott\orcid{0000-0002-6878-9840}\inst{\ref{aff178}}}
										   
\institute{Institute of Cosmology and Gravitation, University of Portsmouth, Portsmouth PO1 3FX, UK\label{aff1}
\and
Fontys University of Applied Sciences, School of ICT, Rachelsmolen 1, 5612 MA Eindhoven, The Netherlands\label{aff2}
\and
University of Applied Sciences and Arts of Northwestern Switzerland, School of Engineering, 5210 Windisch, Switzerland\label{aff3}
\and
Finnish Centre for Astronomy with ESO (FINCA), Quantum, Vesilinnantie 5, University of Turku, 20014 Turku, Finland\label{aff4}
\and
Department of Physics, P.O. Box 64, University of Helsinki, 00014 Helsinki, Finland\label{aff5}
\and
INAF-IASF Milano, Via Alfonso Corti 12, 20133 Milano, Italy\label{aff6}
\and
Dipartimento di Fisica e Scienze della Terra, Universit\`a degli Studi di Ferrara, Via Giuseppe Saragat 1, 44122 Ferrara, Italy\label{aff7}
\and
Departamento F\'isica Aplicada, Universidad Polit\'ecnica de Cartagena, Campus Muralla del Mar, 30202 Cartagena, Murcia, Spain\label{aff8}
\and
Institut de Ci\`{e}ncies del Cosmos (ICCUB), Universitat de Barcelona (IEEC-UB), Mart\'{i} i Franqu\`{e}s 1, 08028 Barcelona, Spain\label{aff9}
\and
Department of Physics, University of Oxford, Keble Road, Oxford OX1 3RH, UK\label{aff10}
\and
University of Applied Sciences and Arts of Northwestern Switzerland, School of Computer Science, 5210 Windisch, Switzerland\label{aff11}
\and
Dipartimento di Fisica e Astronomia "Augusto Righi" - Alma Mater Studiorum Universit\`a di Bologna, via Piero Gobetti 93/2, 40129 Bologna, Italy\label{aff12}
\and
INAF-Osservatorio di Astrofisica e Scienza dello Spazio di Bologna, Via Piero Gobetti 93/3, 40129 Bologna, Italy\label{aff13}
\and
School of Mathematics, Statistics and Physics, Newcastle University, Herschel Building, Newcastle-upon-Tyne, NE1 7RU, UK\label{aff14}
\and
STAR Institute, University of Li{\`e}ge, Quartier Agora, All\'ee du six Ao\^ut 19c, 4000 Li\`ege, Belgium\label{aff15}
\and
Universiteit Gent, Department of Physics and Astronomy, Proeftuinstraat 86 N3, 9000 Ghent, Belgium
\label{aff16}
\and
Instituci\'o Catalana de Recerca i Estudis Avan\c{c}ats (ICREA), Passeig de Llu\'{\i}s Companys 23, 08010 Barcelona, Spain\label{aff17}
\and
Institut de Ciencies de l'Espai (IEEC-CSIC), Campus UAB, Carrer de Can Magrans, s/n Cerdanyola del Vall\'es, 08193 Barcelona, Spain\label{aff18}
\and
School of Physical Sciences, The Open University, Milton Keynes, MK7 6AA, UK\label{aff19}
\and
Jet Propulsion Laboratory, California Institute of Technology, 4800 Oak Grove Drive, Pasadena, CA, 91109, USA\label{aff20}
\and
ESAC/ESA, Camino Bajo del Castillo, s/n., Urb. Villafranca del Castillo, 28692 Villanueva de la Ca\~nada, Madrid, Spain\label{aff21}
\and
INFN-Sezione di Bologna, Viale Berti Pichat 6/2, 40127 Bologna, Italy\label{aff22}
\and
Aix-Marseille Universit\'e, CNRS, CNES, LAM, Marseille, France\label{aff23}
\and
Institut d'Astrophysique de Paris, UMR 7095, CNRS, and Sorbonne Universit\'e, 98 bis boulevard Arago, 75014 Paris, France\label{aff24}
\and
Department of Astronomy, University of Geneva, ch. d'Ecogia 16, 1290 Versoix, Switzerland\label{aff25}
\and
National Centre for Nuclear Research, ul. Pasteura 7, 02-093, Warsaw, Poland\label{aff26}
\and
Laboratoire univers et particules de Montpellier, Universit\'e de Montpellier, CNRS, 34090 Montpellier, France\label{aff27}
\and
Universit\'e de Gen\`eve, D\'epartement de Physique Th\'eorique and Centre for Astroparticle Physics, 24 quai Ernest-Ansermet, CH-1211 Gen\`eve 4, Switzerland\label{aff28}
\and
Universit\'e Paris-Saclay, CNRS, Institut d'astrophysique spatiale, 91405, Orsay, France\label{aff29}
\and
INAF-Osservatorio Astronomico di Brera, Via Brera 28, 20122 Milano, Italy\label{aff30}
\and
IFPU, Institute for Fundamental Physics of the Universe, via Beirut 2, 34151 Trieste, Italy\label{aff31}
\and
INAF-Osservatorio Astronomico di Trieste, Via G. B. Tiepolo 11, 34143 Trieste, Italy\label{aff32}
\and
INFN, Sezione di Trieste, Via Valerio 2, 34127 Trieste TS, Italy\label{aff33}
\and
SISSA, International School for Advanced Studies, Via Bonomea 265, 34136 Trieste TS, Italy\label{aff34}
\and
Dipartimento di Fisica e Astronomia, Universit\`a di Bologna, Via Gobetti 93/2, 40129 Bologna, Italy\label{aff35}
\and
INAF-Osservatorio Astronomico di Padova, Via dell'Osservatorio 5, 35122 Padova, Italy\label{aff36}
\and
Dipartimento di Fisica, Universit\`a di Genova, Via Dodecaneso 33, 16146, Genova, Italy\label{aff37}
\and
INFN-Sezione di Genova, Via Dodecaneso 33, 16146, Genova, Italy\label{aff38}
\and
Department of Physics "E. Pancini", University Federico II, Via Cinthia 6, 80126, Napoli, Italy\label{aff39}
\and
INAF-Osservatorio Astronomico di Capodimonte, Via Moiariello 16, 80131 Napoli, Italy\label{aff40}
\and
Dipartimento di Fisica, Universit\`a degli Studi di Torino, Via P. Giuria 1, 10125 Torino, Italy\label{aff41}
\and
INFN-Sezione di Torino, Via P. Giuria 1, 10125 Torino, Italy\label{aff42}
\and
INAF-Osservatorio Astrofisico di Torino, Via Osservatorio 20, 10025 Pino Torinese (TO), Italy\label{aff43}
\and
Leiden Observatory, Leiden University, Einsteinweg 55, 2333 CC Leiden, The Netherlands\label{aff44}
\and
Centro de Investigaciones Energ\'eticas, Medioambientales y Tecnol\'ogicas (CIEMAT), Avenida Complutense 40, 28040 Madrid, Spain\label{aff45}
\and
Port d'Informaci\'{o} Cient\'{i}fica, Campus UAB, C. Albareda s/n, 08193 Bellaterra (Barcelona), Spain\label{aff46}
\and
Institute of Space Sciences (ICE, CSIC), Campus UAB, Carrer de Can Magrans, s/n, 08193 Barcelona, Spain\label{aff47}
\and
Institut d'Estudis Espacials de Catalunya (IEEC),  Edifici RDIT, Campus UPC, 08860 Castelldefels, Barcelona, Spain\label{aff48}
\and
INAF-Osservatorio Astronomico di Roma, Via Frascati 33, 00078 Monteporzio Catone, Italy\label{aff49}
\and
INFN -- Sezione di Napoli, Via Cinthia 6, 80126, Napoli, Italy\label{aff50}
\and
Dipartimento di Fisica e Astronomia "Augusto Righi" - Alma Mater Studiorum Universit\`a di Bologna, Viale Berti Pichat 6/2, 40127 Bologna, Italy\label{aff51}
\and
Instituto de Astrof\'{\i}sica de Canarias, E-38205 La Laguna, Tenerife, Spain\label{aff52}
\and
Institute for Astronomy, University of Edinburgh, Royal Observatory, Blackford Hill, Edinburgh EH9 3HJ, UK\label{aff53}
\and
European Space Agency/ESRIN, Largo Galileo Galilei 1, 00044 Frascati, Roma, Italy\label{aff54}
\and
Universit\'e Lyon 1, CNRS/IN2P3, IP2I Lyon, Villeurbanne, France\label{aff55}
\and
Universit\'e Lyon 1, CNRS/IN2P3, IUF, IP2I Lyon, Villeurbanne, France \label{aff56}
\and
Mullard Space Science Laboratory, University College London, Holmbury St Mary, Dorking, Surrey RH5 6NT, UK\label{aff57}
\and
Departamento de F\'isica, Faculdade de Ci\^encias, Universidade de Lisboa, Edif\'icio C8, Campo Grande, PT1749-016 Lisboa, Portugal\label{aff58}
\and
Instituto de Astrof\'isica e Ci\^encias do Espa\c{c}o, Faculdade de Ci\^encias, Universidade de Lisboa, Campo Grande, 1749-016 Lisboa, Portugal\label{aff59}
\and
INFN-Padova, Via Marzolo 8, 35131 Padova, Italy\label{aff60}
\and
Aix-Marseille Universit\'e, CNRS/IN2P3, CPPM, Marseille, France\label{aff61}
\and
INAF-Istituto di Astrofisica e Planetologia Spaziali, via del Fosso del Cavaliere, 100, 00100 Roma, Italy\label{aff62}
\and
School of Physics, HH Wills Physics Laboratory, University of Bristol, Tyndall Avenue, Bristol, BS8 1TL, UK\label{aff63}
\and
University Observatory, LMU Faculty of Physics, Scheinerstr.~1, 81679 Munich, Germany\label{aff64}
\and
Max Planck Institute for Extraterrestrial Physics, Giessenbachstr. 1, 85748 Garching, Germany\label{aff65}
\and
Universit\"ats-Sternwarte M\"unchen, Fakult\"at f\"ur Physik, Ludwig-Maximilians-Universit\"at M\"unchen, Scheinerstr.~1, 81679 M\"unchen, Germany\label{aff66}
\and
Institute of Theoretical Astrophysics, University of Oslo, P.O. Box 1029 Blindern, 0315 Oslo, Norway\label{aff67}
\and
Caltech/IPAC, 1200 E. California Blvd., Pasadena, CA 91125, USA\label{aff68}
\and
School of Physics and Astronomy, Lancaster University, Lancaster, LA1 4YB, UK\label{aff69}
\and
Felix Hormuth Engineering, Goethestr. 17, 69181 Leimen, Germany\label{aff70}
\and
Technical University of Denmark, Elektrovej 327, 2800 Kgs. Lyngby, Denmark\label{aff71}
\and
Cosmic Dawn Center (DAWN), Denmark\label{aff72}
\and
Max-Planck-Institut f\"ur Astronomie, K\"onigstuhl 17, 69117 Heidelberg, Germany\label{aff73}
\and
NASA Goddard Space Flight Center, Greenbelt, MD 20771, USA\label{aff74}
\and
Department of Physics and Astronomy, University College London, Gower Street, London WC1E 6BT, UK\label{aff75}
\and
Helsinki Institute of Physics, Gustaf H{\"a}llstr{\"o}min katu 2, University of Helsinki, 00014 Helsinki, Finland\label{aff76}
\and
Laboratoire d'etude de l'Univers et des phenomenes eXtremes, Observatoire de Paris, Universit\'e PSL, Sorbonne Universit\'e, CNRS, 92190 Meudon, France\label{aff77}
\and
SKAO, Jodrell Bank, Lower Withington, Macclesfield SK11 9FT, UK\label{aff78}
\and
Centre de Calcul de l'IN2P3/CNRS, 21 avenue Pierre de Coubertin 69627 Villeurbanne Cedex, France\label{aff79}
\and
Universit\"at Bonn, Argelander-Institut f\"ur Astronomie, Auf dem H\"ugel 71, 53121 Bonn, Germany\label{aff80}
\and
INFN-Sezione di Roma, Piazzale Aldo Moro, 2 - c/o Dipartimento di Fisica, Edificio G. Marconi, 00185 Roma, Italy\label{aff81}
\and
Department of Physics, Institute for Computational Cosmology, Durham University, South Road, Durham, DH1 3LE, UK\label{aff82}
\and
Universit\'e Paris Cit\'e, CNRS, Astroparticule et Cosmologie, 75013 Paris, France\label{aff83}
\and
CNRS-UCB International Research Laboratory, Centre Pierre Bin\'etruy, IRL2007, CPB-IN2P3, Berkeley, USA\label{aff84}
\and
Institute of Physics, Laboratory of Astrophysics, Ecole Polytechnique F\'ed\'erale de Lausanne (EPFL), Observatoire de Sauverny, 1290 Versoix, Switzerland\label{aff85}
\and
Dipartimento di Fisica - Sezione di Astronomia, Universit\`a di Trieste, Via Tiepolo 11, 34131 Trieste, Italy\label{aff86}
\and
Telespazio UK S.L. for European Space Agency (ESA), Camino bajo del Castillo, s/n, Urbanizacion Villafranca del Castillo, Villanueva de la Ca\~nada, 28692 Madrid, Spain\label{aff87}
\and
Institut de F\'{i}sica d'Altes Energies (IFAE), The Barcelona Institute of Science and Technology, Campus UAB, 08193 Bellaterra (Barcelona), Spain\label{aff88}
\and
School of Mathematics and Physics, University of Surrey, Guildford, Surrey, GU2 7XH, UK\label{aff89}
\and
European Space Agency/ESTEC, Keplerlaan 1, 2201 AZ Noordwijk, The Netherlands\label{aff90}
\and
Waterloo Centre for Astrophysics, University of Waterloo, Waterloo, Ontario N2L 3G1, Canada\label{aff91}
\and
Department of Physics and Astronomy, University of Waterloo, Waterloo, Ontario N2L 3G1, Canada\label{aff92}
\and
Perimeter Institute for Theoretical Physics, Waterloo, Ontario N2L 2Y5, Canada\label{aff93}
\and
Universit\'e Paris-Saclay, Universit\'e Paris Cit\'e, CEA, CNRS, AIM, 91191, Gif-sur-Yvette, France\label{aff94}
\and
Space Science Data Center, Italian Space Agency, via del Politecnico snc, 00133 Roma, Italy\label{aff95}
\and
Centre National d'Etudes Spatiales -- Centre spatial de Toulouse, 18 avenue Edouard Belin, 31401 Toulouse Cedex 9, France\label{aff96}
\and
Institute of Space Science, Str. Atomistilor, nr. 409 M\u{a}gurele, Ilfov, 077125, Romania\label{aff97}
\and
Consejo Superior de Investigaciones Cientificas, Calle Serrano 117, 28006 Madrid, Spain\label{aff98}
\and
Universidad de La Laguna, Dpto. Astrof\'\i sica, E-38206 La Laguna, Tenerife, Spain\label{aff99}
\and
Dipartimento di Fisica e Astronomia "G. Galilei", Universit\`a di Padova, Via Marzolo 8, 35131 Padova, Italy\label{aff100}
\and
Instituto de F\'isica Te\'orica UAM-CSIC, Campus de Cantoblanco, 28049 Madrid, Spain\label{aff101}
\and
Institut de Recherche en Astrophysique et Plan\'etologie (IRAP), Universit\'e de Toulouse, CNRS, UPS, CNES, 14 Av. Edouard Belin, 31400 Toulouse, France\label{aff102}
\and
Universit\'e St Joseph; Faculty of Sciences, Beirut, Lebanon\label{aff103}
\and
Departamento de F\'isica, FCFM, Universidad de Chile, Blanco Encalada 2008, Santiago, Chile\label{aff104}
\and
Universit\"at Innsbruck, Institut f\"ur Astro- und Teilchenphysik, Technikerstr. 25/8, 6020 Innsbruck, Austria\label{aff105}
\and
Department of Physics and Helsinki Institute of Physics, Gustaf H\"allstr\"omin katu 2, University of Helsinki, 00014 Helsinki, Finland\label{aff106}
\and
Instituto de Astrof\'isica e Ci\^encias do Espa\c{c}o, Faculdade de Ci\^encias, Universidade de Lisboa, Tapada da Ajuda, 1349-018 Lisboa, Portugal\label{aff107}
\and
Cosmic Dawn Center (DAWN)\label{aff108}
\and
Niels Bohr Institute, University of Copenhagen, Jagtvej 128, 2200 Copenhagen, Denmark\label{aff109}
\and
Universidad Polit\'ecnica de Cartagena, Departamento de Electr\'onica y Tecnolog\'ia de Computadoras,  Plaza del Hospital 1, 30202 Cartagena, Spain\label{aff110}
\and
European University of Technology EUt+, European Union\label{aff111}
\and
Kapteyn Astronomical Institute, University of Groningen, PO Box 800, 9700 AV Groningen, The Netherlands\label{aff112}
\and
Istituto Nazionale di Fisica Nucleare, Sezione di Ferrara, Via Giuseppe Saragat 1, 44122 Ferrara, Italy\label{aff113}
\and
ICSC - Centro Nazionale di Ricerca in High Performance Computing, Big Data e Quantum Computing, Via Magnanelli 2, Bologna, Italy\label{aff114}
\and
INAF, Istituto di Radioastronomia, Via Piero Gobetti 101, 40129 Bologna, Italy\label{aff115}
\and
INFN-Bologna, Via Irnerio 46, 40126 Bologna, Italy\label{aff116}
\and
Astronomical Observatory of the Autonomous Region of the Aosta Valley (OAVdA), Loc. Lignan 39, I-11020, Nus (Aosta Valley), Italy\label{aff117}
\and
Universit\'e C\^{o}te d'Azur, Observatoire de la C\^{o}te d'Azur, CNRS, Laboratoire Lagrange, Bd de l'Observatoire, CS 34229, 06304 Nice cedex 4, France\label{aff118}
\and
Department of Mathematical Physics, Institute of Physics, University of S\~ao Paulo, R. do Mat\~ao 1371, 05508-090, S\~ao Paulo, SP, Brazil\label{aff119}
\and
CEA Saclay, DFR/IRFU, Service d'Astrophysique, Bat. 709, 91191 Gif-sur-Yvette, France\label{aff120}
\and
Oskar Klein Centre for Cosmoparticle Physics, Department of Physics, Stockholm University, Stockholm, SE-106 91, Sweden\label{aff121}
\and
Astrophysics Group, Blackett Laboratory, Imperial College London, London SW7 2AZ, UK\label{aff122}
\and
Dipartimento di Fisica, Sapienza Universit\`a di Roma, Piazzale Aldo Moro 2, 00185 Roma, Italy\label{aff123}
\and
Department of Mathematics and Physics E. De Giorgi, University of Salento, Via per Arnesano, CP-I93, 73100, Lecce, Italy\label{aff124}
\and
INFN, Sezione di Lecce, Via per Arnesano, CP-193, 73100, Lecce, Italy\label{aff125}
\and
INAF-Sezione di Lecce, c/o Dipartimento Matematica e Fisica, Via per Arnesano, 73100, Lecce, Italy\label{aff126}
\and
Institut d'Astrophysique de Paris, 98bis Boulevard Arago, 75014, Paris, France\label{aff127}
\and
ICL, Junia, Universit\'e Catholique de Lille, LITL, 59000 Lille, France\label{aff128}
\and
CERCA/ISO, Department of Physics, Case Western Reserve University, 10900 Euclid Avenue, Cleveland, OH 44106, USA\label{aff129}
\and
Laboratoire Univers et Th\'eorie, Observatoire de Paris, Universit\'e PSL, Universit\'e Paris Cit\'e, CNRS, 92190 Meudon, France\label{aff130}
\and
Dipartimento di Fisica "Aldo Pontremoli", Universit\`a degli Studi di Milano, Via Celoria 16, 20133 Milano, Italy\label{aff131}
\and
INFN-Sezione di Milano, Via Celoria 16, 20133 Milano, Italy\label{aff132}
\and
Departamento de F{\'\i}sica Fundamental. Universidad de Salamanca. Plaza de la Merced s/n. 37008 Salamanca, Spain\label{aff133}
\and
IRFU, CEA, Universit\'e Paris-Saclay 91191 Gif-sur-Yvette Cedex, France\label{aff134}
\and
Universit\'e de Strasbourg, CNRS, Observatoire astronomique de Strasbourg, UMR 7550, 67000 Strasbourg, France\label{aff135}
\and
Center for Data-Driven Discovery, Kavli IPMU (WPI), UTIAS, The University of Tokyo, Kashiwa, Chiba 277-8583, Japan\label{aff136}
\and
Jodrell Bank Centre for Astrophysics, Department of Physics and Astronomy, University of Manchester, Oxford Road, Manchester M13 9PL, UK\label{aff137}
\and
California Institute of Technology, 1200 E California Blvd, Pasadena, CA 91125, USA\label{aff138}
\and
Department of Physics \& Astronomy, University of California Irvine, Irvine CA 92697, USA\label{aff139}
\and
Institute for Astronomy, University of Hawaii, 2680 Woodlawn Drive, Honolulu, HI 96822, USA\label{aff140}
\and
Instituto de F\'isica de Cantabria, Edificio Juan Jord\'a, Avenida de los Castros, 39005 Santander, Spain\label{aff141}
\and
Department of Computer Science, Aalto University, PO Box 15400, Espoo, FI-00 076, Finland\label{aff142}
\and
Ruhr University Bochum, Faculty of Physics and Astronomy, Astronomical Institute (AIRUB), German Centre for Cosmological Lensing (GCCL), 44780 Bochum, Germany\label{aff143}
\and
DARK, Niels Bohr Institute, University of Copenhagen, Jagtvej 155, 2200 Copenhagen, Denmark\label{aff144}
\and
Department of Physics, Centre for Extragalactic Astronomy, Durham University, South Road, Durham, DH1 3LE, UK\label{aff145}
\and
Astrophysics Research Centre, University of KwaZulu-Natal, Westville Campus, Durban 4041, South Africa\label{aff146}
\and
School of Mathematics, Statistics \& Computer Science, University of KwaZulu-Natal, Westville Campus, Durban 4041, South Africa\label{aff147}
\and
Department of Physics and Astronomy, Vesilinnantie 5, University of Turku, 20014 Turku, Finland\label{aff148}
\and
Serco for European Space Agency (ESA), Camino bajo del Castillo, s/n, Urbanizacion Villafranca del Castillo, Villanueva de la Ca\~nada, 28692 Madrid, Spain\label{aff149}
\and
ARC Centre of Excellence for Dark Matter Particle Physics, Melbourne, Australia\label{aff150}
\and
Centre for Astrophysics \& Supercomputing, Swinburne University of Technology,  Hawthorn, Victoria 3122, Australia\label{aff151}
\and
Department of Physics and Astronomy, University of the Western Cape, Bellville, Cape Town, 7535, South Africa\label{aff152}
\and
MTA-CSFK Lend\"ulet Large-Scale Structure Research Group, Konkoly-Thege Mikl\'os \'ut 15-17, H-1121 Budapest, Hungary\label{aff153}
\and
Konkoly Observatory, HUN-REN CSFK, MTA Centre of Excellence, Budapest, Konkoly Thege Mikl\'os {\'u}t 15-17. H-1121, Hungary\label{aff154}
\and
Brazilian Center for Research in Physics (CBPF), Dr. Xavier Sigaud st. 150, zip 22290-180, Rio de Janeiro, RJ, Brazil\label{aff155}
\and
Institute for Theoretical Particle Physics and Cosmology (TTK), RWTH Aachen University, 52056 Aachen, Germany\label{aff156}
\and
INAF-Osservatorio Astrofisico di Arcetri, Largo E. Fermi 5, 50125, Firenze, Italy\label{aff157}
\and
Centro de Astrof\'{\i}sica da Universidade do Porto, Rua das Estrelas, 4150-762 Porto, Portugal\label{aff158}
\and
Instituto de Astrof\'isica e Ci\^encias do Espa\c{c}o, Universidade do Porto, CAUP, Rua das Estrelas, PT4150-762 Porto, Portugal\label{aff159}
\and
HE Space for European Space Agency (ESA), Camino bajo del Castillo, s/n, Urbanizacion Villafranca del Castillo, Villanueva de la Ca\~nada, 28692 Madrid, Spain\label{aff160}
\and
Center for Frontier Science, Chiba University, 1-33 Yayoi-cho, Inage-ku, Chiba 263-8522, Japan\label{aff161}
\and
Department of Physics, Graduate School of Science, Chiba University, 1-33 Yayoi-Cho, Inage-Ku, Chiba 263-8522, Japan\label{aff162}
\and
Department of Astrophysics, University of Zurich, Winterthurerstrasse 190, 8057 Zurich, Switzerland\label{aff163}
\and
INAF - Osservatorio Astronomico d'Abruzzo, Via Maggini, 64100, Teramo, Italy\label{aff164}
\and
Theoretical astrophysics, Department of Physics and Astronomy, Uppsala University, Box 524, 751 20 Uppsala, Sweden\label{aff165}
\and
Mathematical Institute, University of Leiden, Einsteinweg 55, 2333 CA Leiden, The Netherlands\label{aff166}
\and
Institute of Astronomy, University of Cambridge, Madingley Road, Cambridge CB3 0HA, UK\label{aff167}
\and
Univ. Lille, CNRS, Centrale Lille, UMR 9189 CRIStAL, 59000 Lille, France\label{aff168}
\and
Center for Astrophysics and Cosmology, University of Nova Gorica, Nova Gorica, Slovenia\label{aff169}
\and
Department of Astrophysical Sciences, Peyton Hall, Princeton University, Princeton, NJ 08544, USA\label{aff170}
\and
Fakult\"at f\"ur Physik, Universit\"at Bielefeld, Postfach 100131, 33501 Bielefeld, Germany\label{aff171}
\and
Space physics and astronomy research unit, University of Oulu, Pentti Kaiteran katu 1, FI-90014 Oulu, Finland\label{aff172}
\and
Department of Physics and Astronomy, Lehman College of the CUNY, Bronx, NY 10468, USA\label{aff173}
\and
American Museum of Natural History, Department of Astrophysics, New York, NY 10024, USA\label{aff174}
\and
International Centre for Theoretical Physics (ICTP), Strada Costiera 11, 34151 Trieste, Italy\label{aff175}
\and
Center for Computational Astrophysics, Flatiron Institute, 162 5th Avenue, 10010, New York, NY, USA\label{aff176}
\and
SRON Netherlands Institute for Space Research, Landleven 12, 9747 AD, Groningen, The Netherlands\label{aff177}
\and
Department of Physics and Astronomy, University of British Columbia, Vancouver, BC V6T 1Z1, Canada\label{aff178}}    
   
%
%
%
%


%
%

\abstract{
The Euclid Data Release 1 (DR1) strong lens search presents ECLIPSE, the largest catalogue of strong gravitational lens candidates to date, with $>$15\,000 systems identified across 1900~deg$^2$ of sky. It was constructed using five neural networks applied to approximately 68 million galaxies, followed by citizen science and expert visual inspection. In this work, we compare how the independently developed networks use image information to identify their selection of strong lens candidates. The search combines convolutional neural networks and a vision transformer, using the same pool of data but with different image processing and training choices. We investigate why these networks find different systems and what that means for the completeness of our catalogue. We evaluate all networks on a 50~deg$^2$ calibration set and their application to the full DR1 sample. We find that the networks score candidates very differently, and that combining them recovers over 500 lens candidates within our calibration set at an inspection budget of 2000 objects. On the full DR1 sample, we show that individual networks have distinct biases: some networks favour systems with a larger Einstein radius while others are more sensitive to compact configurations. We also investigate network preferences in terms of lens light profile, mass distribution geometry, photometric redshift, and magnitudes. These findings provide a detailed analysis of how neural network selection biases have shaped the Euclid DR1 ECLIPSE catalogue. The results inform how networks should be combined in future \Euclid data releases and upcoming wide-field surveys such as LSST or \textit{Roman} to maximise both completeness and diversity of the recovered lens population.
}

%
%

\titlerunning{Learning machine learning for Euclid DR1}
\authorrunning{Euclid Collaboration: Vincken et al.}
%
%
   
\maketitle
%
%
%
%

\section{\label{sc:Intro}Introduction}
Strong gravitational lensing is a phenomenon in our universe where the mass of a foreground galaxy visibly distorts the light of a background source. Understanding these rare systems broadens our capability to study mass distributions in galaxies \citep[e.g.,][]{Gavazzi_2007, Nightingale_2019,Sonnenfeld2024,Shajib_2024}, observe distant sources \citep{Welch2022}, understand core cosmological parameters such as the Hubble constant \citep{Wong2019, TDCOSMO2025}, and the potential properties of dark matter and dark energy \citep{vegetti2023stronggravitationallensingprobe,li2024}. 

The current known strong lens candidates have been identified using a number of methods, from serendipitous discoveries \citep[e.g.,][]{Walsh1979}, feature extraction methods \citep[e.g.,][]{alard2006automateddetectiongravitationalarcs,Gavazzi_2014,Joseph_2014}, to more modern techniques such as deep learning using neural networks \citep[NN; e.g.,][]{Q1-SP053, storfer2026}. Using these methods across many individual surveys has increased the number of known strong lens candidates from hundreds to thousands over the recent years \citep[e.g.,][]{Jacobs_2017,Petrillo_2017,Huang_2021}. The \Euclid mission will deliver the first survey with the volume, depth, and image quality to deliver over \num{100000} lenses.

With the Quick Data Release 1 (Q1), \Euclid released its first catalogue of strong lens candidates \citep{Q1-SP048, Q1-Ecker}, discovered by a combined approach of machine learning \citep[ML,][]{Q1-SP053, Q1-Vincken} and visual inspection (VI) campaigns \citep{Q1-SP052, Q1-SP059}. These efforts led to about 500 high-confidence grade~A and grade~B lens candidates by searching through $63.1$~deg$^2$ of sky targeting a selection of approximately 1.1 million galaxies.

Building on Q1, for the Euclid Data Release 1 (DR1; \citealp{DR1cite}), we search for strong lens systems in an area of $\sim$1900~deg$^2$. To build the DR1 strong lens catalogue, we use five independently developed NN to score and rank all objects. While the use of multiple NN is already established \citep[e.g.,][]{Savary2022UNIONS, AndikaQSO2023, Schuldt2025HOLISMOKES, gonzalez2025doesmachinelearningwork, Herle2024}, with the DR1 search now operating at a much larger scale, understanding the behaviour of these networks becomes more important. Each network differs in architecture, training data input, and training strategy, which means they may be sensitive to different types of lens systems. While the Q1 search already showed that combining networks recovers more candidates than any individual network, it remains unclear what drives these differences. Do the networks find different systems because they respond to different physical properties of the lens, such as Einstein radius, galaxy morphology, or brightness? And if so, does our combination of networks introduce biases into the final catalogue by over-representing certain types of systems?

In this paper we address these questions by analysing the behaviour of networks used for the DR1 search. We evaluate the networks on a dedicated 50~deg$^2$ calibration set and apply them to the full DR1 sample, followed by the visual inspection pipeline presented by \citet{DR1-SL02}. This search yielded a catalogue of $>$\num{15000} strong lens candidates as presented by \citet{DR1-SL01}, increasing the number of known candidates by more than an order of magnitude compared to Q1. Alongside the catalogue, lens model fits are presented by \citet{DR1-SL03} and photometric redshifts by \citet{DR1-SL04}. Furthermore, \cite{DR1-SL08} presents a set of double-source plane lenses, \Ecker{} investigates the population of edge-on disk lenses, \cite{DR1-SL06} presents an analysis on impact of image processing for neural networks, and \citet{DR1-SL07} provides a new method for automated creation of segmentation maps for strong lens components. Using these outputs, we investigate whether individual networks show preferences for particular lens properties, where biases lie, and whether the networks are complementary in providing a complete selection for our catalogue.

In Sect.~\ref{sc:Data} we describe the training data, inference set, and the DR1 calibration set. The methodology used to compare the networks is presented in Sect.~\ref{sc:Method}, followed by a description of the participating networks in Sect.~\ref{sc:ML}. Results on the calibration set, including ranking consensus, union recovery, and performance metrics, are given in Sect.~\ref{sc:ML-behaviour}. We characterise the network biases on the full DR1 sample in Sect.~\ref{sc:Characterisation} and discuss the implications for catalogue construction in Sect.~\ref{sc:DiscussionConclusion}.

\section{\label{sc:Data}Data}
All participating networks have access to a common data pool, consisting of both simulated lens systems and real \Euclid observations. We describe each in turn.

\subsection{Simulated training samples}
The synthetic training data consist of real deflector galaxies observed by \Euclid with lensed background sources. These simulations are made by two independent approaches, hereafter called set S1 and S2, described in detail by \cite{Q1-SP053} and \cite{Q1-Vincken}.

S1 simulations \citep{Rojas2022, Q1-SP052} use the {\tt Lenstronomy}\footnote{\url{https://github.com/lenstronomy/lenstronomy}} package \citep{Birrer_2018, Birrer_2021} with Luminous Red Galaxy (LRG) deflectors and real \textit{Hubble} Space Telescope (HST) images as background sources, lensed through an single isothermal ellipsoid (SIE) mass model. This set consists of approximately \num{11000} systems.

S2 simulations use the \texttt{GLAMER} lensing package \citep{Metcalf_2014} with deflectors matched to the Flagship simulation \citep{EuclidSkyFlagship} and simulated sources based on S\'ersic components drawn from HST Ultra-Deep Field galaxies. This set consists of about 5000 systems. Both sets were downsampled to \Euclid pixel scales, PSF-convolved, and scaled to match \Euclid image properties.

\subsection{Q1 and DR1 training sets}
Labelled observational data were available from two \Euclid releases. All samples used a similar grading system as described in detail by \cite{AcevedoBarroso24}. In this grading system, grade~A corresponds to high-confidence lens systems showing clear lensing features, grade~B to systems showing lensing features but where additional information is required to confirm, grade~C to low-confidence systems showing features that can be explained without resorting to gravitational lensing, and grade~X to objects flagged by ML networks but classified as non-lenses by experts.

Besides the simulation sets, the primary data source we use for training is from the Euclid Q1 Strong Lensing Discovery Engine (SLDE) catalogue presented in \cite{Q1-SP048}. The creation of this catalogue began with a VI search by \cite{Q1-SP052} that identified 38 grade~A and 40 grade~B lens candidates. This was followed by the main ML search by \cite{Q1-SP053}, which provided 250 grade~A, 247 grade~B, and 585 grade~C labelled systems. Three additional searches later extended this catalogue. \cite{Q1-Ecker} contributed 38 grade~A and 34 grade~B candidates by improving the Q1 selection function. \cite{Q1-Vincken} added eight grade~A, 26 grade~B, and 65 grade~C candidates through a follow-up search of the Q1 data to recover missed candidates. \cite{Q1-Xu} contributed nine grade~A and 121 grade~B candidates from an independent search. All of these searches were supported by VI campaigns using the Space Warps (SW) and Galaxy Judges (GJ) platform, where citizen science volunteers and strong lensing experts classify potential lenses through visually inspecting object cutouts \citep{Q1-SP059}.

Besides the main Q1 strong lens search, a morphology search by \cite{Q1-SP047} identified 56 grade~A and 19 grade~B systems. Another three grade~A lens candidates were discovered serendipitously by members of the Euclid Consortium (EC).

An internal search by the EC Strong Lensing Science Working Group (SL-SWG) within \num{200}\,deg$^2$ of DR1 data yielded 791 grade~A, 321 grade~B, and 309 grade~C candidates. This search was conducted ahead of the full DR1 effort to expand the set of labelled real lenses available for retraining, as larger and more diverse training samples improve network performance. Hereafter this search is referred to as the \num{200}\,deg$^2$ rapid search and described in detail by \citet{DR1-SL02}. All counts of graded samples across the \Euclid searches are summarised in Table~\ref{tab:data_summary}.

\begin{table*}
\centering
\setlength{\tabcolsep}{8pt}
\caption{Summary of labelled strong lens candidates available for training from Euclid Q1, DR1, and the 50~deg$^2$ calibration set, each from individual searches not accounting for overlap.}
\label{tab:data_summary}
\begin{tabular}{llrrrr}
\hline\hline
Source & Discovery & A & B & C & X \\
\hline
\multicolumn{6}{l}{\textit{Euclid ERO}} \\
\hspace{1em}\cite{AcevedoBarroso24} & VI & 3 & 13 & 52 &  \ldots\\
\hline
\multicolumn{6}{l}{\textit{Euclid Q1}} \\
\hspace{1em}\cite{Q1-SP052}& VI  & 38 & 40 & \ldots &  \ldots\\
\hspace{1em}\cite{Q1-SP048} & ML/VI & 250 & 247 & 585 & 6280 \\
\hspace{1em}\cite{Q1-Ecker} & ML/VI & 38 & 34 & 51 & \ldots \\
\hspace{1em}\cite{Q1-Vincken} & ML/VI & 8 & 26 & 65 & 808 \\
\hspace{1em}\cite{Q1-Xu} & ML/VI & 9 & 121 & \ldots & \ldots \\
\hspace{1em}\cite{Q1-SP047} & ML/VI & 56 & 19 & \ldots & \ldots \\
\hspace{1em}EC members & Serendipitous & 3 & \ldots & \ldots & \ldots \\
\hline
\multicolumn{6}{l}{\textit{Euclid DR1}} \\
\hspace{1em}\num{200}\,deg$^2$ rapid search & ML/VI & 791 & 321 & 309 & 825 \\
\hline\hline
\multicolumn{6}{l}{\textit{Euclid DR1 calibration set}} \\
\hspace{1em}50\,deg$^2$ (raw) & ML/VI & 334 & 293 & 377 & 2111 \\
\hspace{1em}50\,deg$^2$ (deduplicated) & ML/VI & 251 & 249 & 336 & 2075 \\
\hline
\label{LastPage}
\end{tabular}
\tablefoot{Grades indicate candidate quality: A (definite lens), B (probable lens), C (possible lens), and X (rejected as non-lens). Discovery methods include visual inspection (VI), machine learning (ML), and serendipitous discovery. The calibration set comprises candidates from a 50~deg$^2$ region within DR1, with raw counts showing all candidates and deduplicated counts removing any duplicate systems. Grade~X objects serve as non-lens contaminants for training.}
\end{table*}

\begin{table}
\centering
\caption{Summary of labelled non-lens examples available for training.}
\label{tab:nonlens_summary}
\begin{tabular}{llr}
\hline\hline
Discovery & Type & Count \\
\hline
\hspace{1em}VI $^{[1]}$ & Spirals & 2578 \\
\hspace{1em}VI $^{[1]}$ & Mergers & 250 \\
\hspace{1em}VI $^{[1]}$ & Ring galaxies & 61 \\
\hspace{1em}VI $^{[1]}$ & Other non-lens galaxies & 2477 \\
\hspace{1em}S1 footprint $^{[1]}$ & LRGs & 2700 \\
\hspace{1em}S2 footprint & Random galaxies & 4000 \\
\hline
\end{tabular}
\smallskip

\textbf{References:}
\hspace{1em}[1] \cite{Q1-SP052}
\end{table}

\subsubsection{\label{sc:TestSet}DR1 calibration set}
A dedicated 50\,deg$^2$ region of the DR1 footprint is withheld from all training and used exclusively for evaluation of all participating ML networks. A calibration set of this size is needed because strong lenses are rare, and a smaller area would contain too few systems to reliably compare the networks and analyse their capabilities. The results of this validation are also used to determine the VI strategy for the entire DR1 sample. 

To create the calibration set, the five participating networks collectively scored \num{1384127} objects across the 50\,deg$^2$ field. Each object was assigned a lens likelihood score between 0 and 1, and was then ranked from highest to lowest scored object. The top ranked candidates were passed to a citizen science classification round on the SW platform. To balance completeness against the volume of inspection budgets, the objects shown were weighted by network rank:  all of the top 2000 objects from each network were included, 50\% of objects ranked between 2000 and 6000, and 25\% of objects ranked between 6000 and \num{12000}. In addition, $\sim$\num{30000} randomly selected objects were shown as a control sample. This sample is retrieved from the parent selection following the criteria described by \cite{DR1-SL02}.

Candidates that passed the SW round proceeded to expert validation on the GJ platform, where each system was assigned one of the four grades. The expert validation of this set yielded 334 grade~A, 293 grade~B, 377 grade~C, and 2111 grade~X classifications. After de-duplication of off-centered objects, these reduce to 251 grade~A, 249 grade~B, 336 grade~C, and 2075 grade~X respectively.

NN can be sensitive to partial arcs appearing in the corners of cutouts, which can lead to the same physical system being scored highly in two separate cutouts centered on different features. Even though they are duplicate systems, this does not constitute an error on the network's part, and collapsing such cases into a single entry would unfairly penalize networks with this capability. Therefore, the ML analysis in this work uses the raw counts.

\subsubsection{\label{sc:InferenceSet}DR1 inference set}
The full DR1 inference set is constructed from the DR1 object catalogue (MER; \citealt{DR1-TP009}). This catalogue contains approximately 900 million detected sources \citep{DR1cite} across $\sim$1900~deg$^2$ of sky.

Since only a small fraction of these objects are expected to be suitable for our search, filtering is applied to identify observations that have detectable lensing features. The criteria of this selection function are described in detail by \citet{DR1-SL01}, which includes requirements on magnitude, surface brightness, and compactness. After applying these criteria, approximately 68 million objects remain. This forms the inference dataset used for the full DR1 strong lens search, separate from the dedicated 50\,deg$^2$ calibration set and excluding objects within the Q1 and deep field region.

\subsubsection{Non-lens contaminants set}
The non-lens dataset consists of several sets of labelled non-lenses, including grade~X objects from the Q1 and DR1 searches. These add a complementary source of contaminants: objects that were flagged by at least one ML network but rejected during VI. These total to 6280 systems from main Q1 search, another 808 systems from the follow-up searches, and 825 from the internal 200\,deg$^2$ DR1 search conducted by the \Euclid SL-SWG.

Additionally, \cite{Q1-SP052} provided 2578 spirals, 250 mergers, 61 ring galaxies, 2477 general non-lens galaxies, and 2700 LRG, the latter also forming the deflector base for the S1 simulations. Spirals, mergers, and rings are of particular importance as they are known to produce common false-positives in lens searches \citep{Rojas2022}. A further set of approximately 4000 randomly selected non-lens galaxies from the S2 simulation footprint are also available. The summary of these systems is given in Table~\ref{tab:nonlens_summary}. The grade~X numbers are summarized together with the other grades in Table~\ref{tab:data_summary}.

\subsection{Image formats and pixel scaling}
All data sets described in this work are available in four \Euclid bands: the single broad optical band of the VIS instrument \citep[\IE, 0.55--0.95\,$\mu$m,][]{EuclidSkyVIS} and the three bands of the Near Infrared Spectrometer and Photometer (NISP) instrument \citep[\YE: 949.6--1212.3 nm; \JE: 1167.6--1567.0 nm; \HE: 1521.5--2021.4 nm,][]{EuclidSkyNISP}.

Every network transforms and augments the pixel values using their individual approaches, but they all make use of arcsinh scaling. This follows the finding of earlier Q1 strong lensing searches, where arcsinh-scaled inputs yield better network performance than other scaling methods \citep{Q1-SP053, Q1-Vincken}.

Finally, all data sets are available in two image formats. The first is FITS (Flexible Image Transport System), where files contain all data from the telescope, including information of every photometric band, and the full pixel data suitable for any processing pipeline. The second format provided is JPEG (Joint Photographic Experts Group), which is used to reduce I/O (input/output) overhead at the scale of millions of objects. JPEG is a compressed image format with limited dynamic range, whereas FITS retains the full precision of the original observations. However, using this compression reduces the total data set size by an order of magnitude compared to the multi-band FITS format. 

\section{\label{sc:Method}Methodology}
All ML networks in this study address a binary classification problem: distinguishing strong gravitational lens systems from non-lens contaminants. Each network outputs a continuous score between 0 and 1 for every inferred object, representing the predicted probability score that the object is a lens. Objects are then ranked by these scores, with higher scores indicating higher lens likelihood.

\subsection{\label{sc:performance}Performance metrics}
We evaluate network performance using receiver operating characteristic (ROC) curves and precision-recall (PR) curves on the DR1 calibration set. The ROC curves go hand in hand with the area under the curve (AUC) which summarizes overall ranking quality, with AUC = 1 indicating perfect separation and AUC = 0.5 indicating random ranking.

For these metrics, a positive class is defined as grade~A and grade~B systems (true lenses) and a negative class consists of grade X systems (non-lenses). We define $N_{\text{TP}}$ as the number of positives correctly identified as lenses, $N_{\text{FP}}$ as the number of negatives incorrectly identified as lenses, $N_{\text{TN}}$ as the number of negatives correctly identified as non-lenses, and $N_{\text{FN}}$ as the number of positives incorrectly identified as non-lenses.

The ROC curve evaluates performance across all possible classification thresholds by plotting the true positive rate against the false positive rate. These rates are defined as
\begin{equation}
\text{TPR} = \frac{N_{\text{TP}}}{N_{\text{TP}} + N_{\text{FN}}}, \quad \text{FPR} = \frac{N_{\text{FP}}}{N_{\text{FP}} + N_{\text{TN}}}~,
\end{equation}
\noindent where TPR measures the fraction of true lenses recovered and FPR measures the fraction of non-lenses incorrectly flagged. 

The PR curve evaluates precision against recall across all possible thresholds, defined as

\begin{equation}
\text{Precision} = \frac{N_{\text{TP}}}{N_{\text{TP}} + N_{\text{FP}}}, \quad \text{Recall} = \frac{N_{\text{TP}}}{N_{\text{TP}} + N_{\text{FN}}}~,
\end{equation}
where precision measures the purity of the selected sample (fraction of flagged objects that are true lenses) and recall is equivalent to TPR (fraction of true lenses recovered). In astronomy, precision and recall are commonly referred to as reliability and completeness, respectively. In highly imbalanced scenarios like ours where lenses are extremely rare, the PR curve captures the purity of top-ranked samples, which is relevant for limited VI budgets.

\subsection{\label{sc:spearman-rank}Ranking consensus}
To quantify the agreement between different networks we compute the Spearman rank correlation coefficient $r_\sfont{S}$ between their predicted scores. Since multiple objects can receive the same score from a network, we use the tie-corrected formulation as given in \cite{ZarZ005}, which assigns tied objects the average of the ranks they would have occupied and adjusts the correlation accordingly.

Values range from $-1$ to $+1$, where $r_\sfont{S} = 1$ indicates perfect agreement, $r_\sfont{S} = 0$ indicates no correlation, and $r_\sfont{S} = -1$ indicates complete disagreement. High correlation suggests networks identify similar candidates while low correlation suggests complementary behaviour. We compute pairwise correlations across all possible NN combinations.

\subsection{\label{sc:LOO}Individual network contribution}
To assess the unique contribution of each network, we perform a leave-one-out analysis where each network is excluded from the union and the decrease in recovered grade~A and grade~B systems is measured. For each network $i$, $N_{\text{all}}$ is the number of grade~A and grade~B systems recovered by the union of all networks' top-$N$ candidates, and $N_{-i}$ is the number recovered when network $i$ is excluded. The unique contribution of network $i$ is

\begin{equation}
\Delta_i = N_{\text{all}} - N_{-i}~,
\end{equation}
where a large $\Delta_i$ indicates that network $i$ recovers systems missed by all other networks, while $\Delta_i = 0$ indicates complete redundancy.

\subsection{\label{sc:combination}Recovery from network unions}
While the Spearman correlation quantifies the similarity of rankings between two networks, it does not directly tell us how many lens candidates would be recovered when combining multiple networks. To assess this, we measure the recovery of grade~A and grade~B systems as a function of the number of networks combined. For each combination size $k$ (from one to the total number of networks), we take the top-$N$ ranked objects from each of the $k$ selected networks and form the union of their candidates. We then count the number of grade~A and grade~B systems present in this union. By computing this for every possible combination of $k$ networks, we obtain a distribution of recovered systems for each combination size.

\subsection{\label{sc:einstein}Einstein radius}
To assess if our networks show biases to the size of systems they are selecting, we measure the Einstein radius ($\theta_\sfont{E}$) for all grade~A and grade~B systems. The measurements are obtained from the \texttt{PyAutoLens} lens modelling pipeline \citep{Nightingale2018, Nightingale2021}, applied to the ECLIPSE catalogue as described by \citet{DR1-SL03}. 

We then take the global rank $R_{i,j}$ of object $j$ for network $i$ as its position when all objects are ordered by decreasing score,
\begin{equation}
R_{i,j} = \operatorname{rank}(s_{i,j})~,
\end{equation}
where $s_{i,j}$ is the score assigned to object $j$ by network $i$, with rank 1 corresponding to the highest score. For each network, we calculate the median of $R_{i,j}$ for the recovered grade~A and grade~B systems as a function of $\theta_\sfont{E}$. A network biased toward larger systems will recover high $\theta_\sfont{E}$ systems in its top-$N$ ranked candidates.

\subsection{Photometric redshift}
The photometric redshift (photo-$z$) of the lens estimates how much its light is shifted toward longer wavelengths. At higher photo-$z$, the foreground lens galaxy may appear fainter, which can affect how easily the lensing system is identified. We use the photo-$z$ of the lens from \citet{DR1-SL04}, and plot the values against the median of $R_{i,j}$ to identify whether the networks recover lenses differently across redshift.

\subsection{\label{sc:sersic}Lens S\'ersic index}
The S\'ersic index $n$ of the lens galaxy describes its light concentration, with higher values corresponding to early-type galaxies and lower values to disk galaxies. Most known strong lenses are early-type galaxies and many searches have preferred to target these systems \citep{Sonnenfeld_2023}. This is because the lensing cross section is much higher for early-type galaxies than for other galaxies (e.g., spirals), so the imbalance is expected and reflects the characteristics of the typical lens population. As a result, training sets are often dominated by early-type lenses, and other morphologies may be under-represented.

Using S\'ersic index measurements from the \texttt{PyAutoLens} light profile fits, we plot the median of $R_{i,j}$ as a function of S\'ersic index. This reveals whether networks preferentially recover high S\'ersic index (early-type) lenses over lower S\'ersic index (disk-dominated) systems.

\subsection{\label{sc:magnitude}Lens and source magnitudes}
The magnitudes of the lens and source galaxy affect how easily a strong lensing system can be identified. Brighter systems produce higher signal-to-noise ratio (S/N) features, which may make them easier for networks to detect. 

We examine the \IE-band magnitude of the source and the lens separately, as well as the total magnitude obtained by summing both fluxes. The total magnitude represents the overall brightness of the system as seen by the network before any decomposition. Separating the source and lens contributions allows us to distinguish whether a network preference is driven by the brightness of the lensing features themselves or by the brightness of the foreground galaxy. 

The source and lens fluxes are taken from the model fits, where the lens flux corresponds to the total flux of the foreground S\'ersic model and the source flux to the total lensed S\'ersic source flux in the image plane, summed over all multiple images. These are observed magnitudes and are not corrected for lensing magnification. The flux densities are in units of $\mu$Jy and are converted to AB magnitudes. We plot these magnitudes against the median of $R_{i,j}$ to identify whether the networks prefer higher or lower magnitude systems.

\subsection{\label{sc:ellipticity}Mass distribution geometry}
The ellipticity of the mass distribution affects the number and shape of lensed images. We use the axis ratio $q = b/a$ of the mass profile, where $b$ and $a$ are the minor and major axes, derived from the \texttt{PyAutoLens} lens modelling outputs. A value of $q = 1$ corresponds to a circular mass distribution and lower values indicate increasing ellipticity.

However, besides the mass ellipticity, external shear also affects the lensing configuration. External shear is a distortion of lensed images beyond that produced by the ellipticity of the mass profile, and can include contributions from nearby galaxies or large-scale structures. A spherical mass distribution with large shear can produce similar image configurations to an elliptical mass distribution with little to no shear. To capture the full dependence on mass model geometry, we therefore also examine the external shear from the lens model fits. The lens model provides two shear components $\gamma_1^{\rm ext}$ and $\gamma_2^{\rm ext}$, following \cite{DR1-SL03}, we convert to the shear magnitude

\begin{equation}
\gamma^{\rm ext} = \sqrt{(\gamma_1^{\rm ext})^2 + (\gamma_2^{\rm ext})^2}~.
\end{equation}
For each network, we plot the axis ratio and shear magnitude against the median of $R_{i,j}$. A network biased toward rounder or more elliptical systems will show a trend in $q$ with recovery order, and similarly any preference for stronger or weaker shear will be visible.

\section{\label{sc:ML}Participating networks}
Five ML networks were created for our search for strong lensing candidates. Each network differs in its use of photometric bands, image format, pre-training strategy, and early stopping criterion (i.e.\ the metric used to decide when training has converged), all of which can influence ranking performance. Although every network was independently developed and trained, they all used the same common data pool described in Sect.~\ref{sc:Data}. 

The networks fall into two architectural families. Convolutional neural networks (CNNs; \citealp{LeCun89}) process images through layers of local filters, each combining information from a small spatial neighbourhood. The Vision Transformer (ViT; \citealp{dosovitskiy2021imageworth16x16words}) instead divides the image into fixed-size patches and relates all patches to one another, giving it access to the full spatial extent of the image from the start. In turn we describe the key training and architecture properties of each network, with their main aspects summarised in Table~\ref{tab:ml_summary}.

\begin{table*}
\centering
\setlength{\tabcolsep}{8pt}
\caption{Summary of the five ML networks used in the DR1 strong lens search.}
\label{tab:ml_summary}
\begin{tabular}{llllllll}
\hline\hline
Network & Architecture & Bands & Input format & Input scaling & Pre-trained & Early stopping metric \\
\hline
NN1 & ViT & \IE  & JPEG & A & Yes & AUC \\
NN2 & CNN & \IE & JPEG & A &  Yes & Val. loss \\
NN3 & CNN & \IE & JPEG & A & No & \ldots \\
NN4 & CNN & \IE  & JPEG & A  & Yes & Val. loss \\
NN5 & CNN & \IE + \JE + \HE & FITS & A/S/C  & Yes & PR AUC \\
\hline
\end{tabular}
\tablefoot{Networks use either the VIS \IE-band alone or in combination with NISP bands (\YE, \JE, \HE). Input scaling methods include arcsinh (A), Sobel edge detection (S), and CLAHE contrast enhancement (C). Images were used in either FITS or JPEG format. Pre-trained indicates whether the network was initialised from a pre-trained network or trained entirely from scratch. Early stopping metric is given where applicable.}
\end{table*}

\begin{figure*}
    \includegraphics[width=\textwidth]{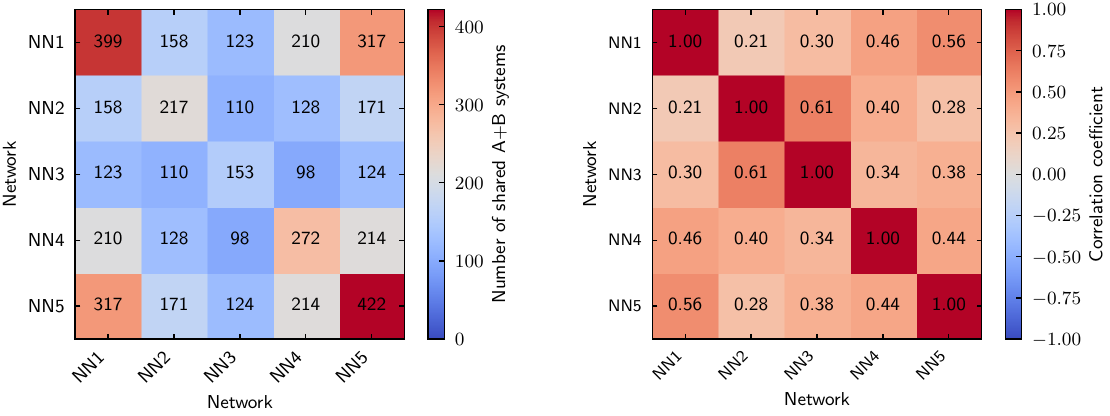}
    \caption{Pairwise comparison of the five networks. \textit{Left}: the number of grade~A+B systems recovered by both networks in each pair within the top \num{10000} highest-scored candidates; diagonal entries show the total number recovered by each individual network. The colour gradient runs from low overlap (blue) to high overlap (red). \textit{Right}: Spearman rank correlation between the predicted scores of each pair of networks, computed over all grade~A and grade~B systems in our calibration set. The colour gradient runs from negative correlation (blue), through no correlation (white), to perfect correlation (red). In both panels, the $x$- and $y$-axes list the five networks.}
    \label{fig:ranking-agreement}
\end{figure*}

\begin{figure*}
    \centering
    \includegraphics[width=\textwidth]{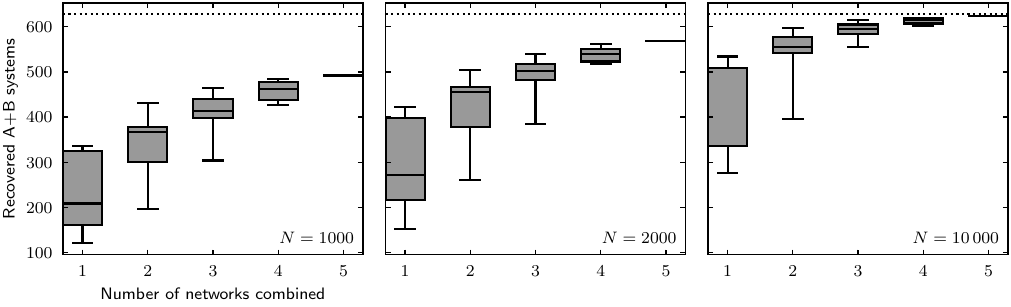}
    \caption{Recovery of grade~A and grade~B candidates when combining multiple networks, shown for inspection thresholds of top 1000 (left), 2000 (centre), and 10\,000 (right) candidates. The $x$-axis shows the number of networks combined and the $y$-axis shows the number of recovered grade~A and grade~B systems. Each of the five networks independently ranks images by predicted lens probability; the recovered candidates are taken as the union of the top-$N$ lists across all combinations of $k$ networks. For each union size $k$, the box spans the middle 50\% of recovered counts across all combinations of that size, with the central line marking the median and whiskers extending to the best and worst performing combination. The dotted line marks the total number of grade~A and grade~B systems in the calibration set.}
    \label{fig:combination_recovery}
\end{figure*}

\begin{figure}
    \centering
    \includegraphics[width=\linewidth]{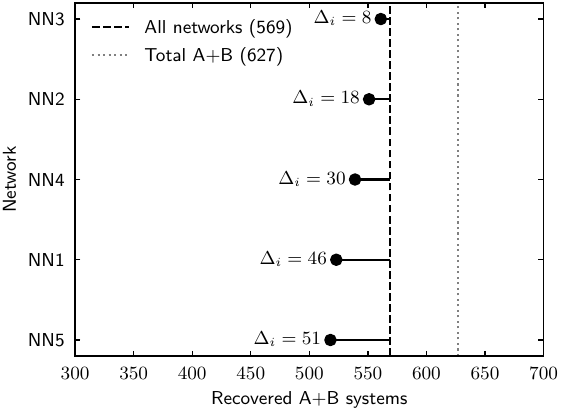}
    \caption{Leave-one-out analysis on the recovery of grade~A and grade~B lens candidates for the top 2000 highest-scored candidates per network. The $x$-axis shows the number of recovered grade~A and grade~B systems and the $y$-axis lists the excluded network. Each dot marks the number of systems recovered by the union of the remaining four networks. The horizontal line spans from this value to the full five-network recovery (dashed line). The annotation $\Delta_i$ gives the unique contribution of the excluded network, defined as the difference between the full five-network recovery and the four-network recovery. Networks are ordered from smallest to largest $\Delta_i$. The dotted line marks the total number of grade~A and grade~B systems in the calibration set.}
    \label{fig:leave_one_out}
\end{figure}

\begin{figure*}
    \centering
    \includegraphics[width=\textwidth]{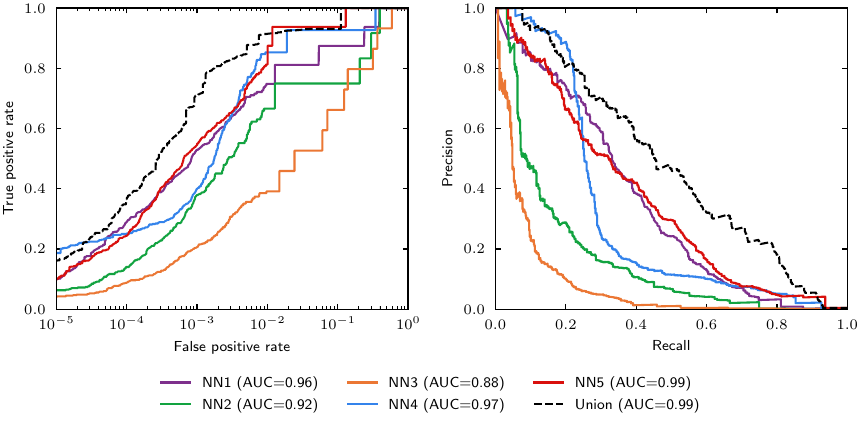}
    \caption{Performance comparison of the five ML networks on the 50~deg$^2$ calibration set. \textit{Left}: ROC curves showing true positive rate on the $y$-axis versus false positive rate on the $x$-axis (logarithmic scale). The area under the curve (AUC) is shown in the legend. \textit{Right}: PR curves showing precision on the $y$-axis versus recall on the $x$-axis. The union of all networks combines predictions from all five networks by averaging their scores.}
    \label{fig:roc_pr_combined}
\end{figure*}

\begin{figure*}
\centering
\includegraphics[width=\linewidth]{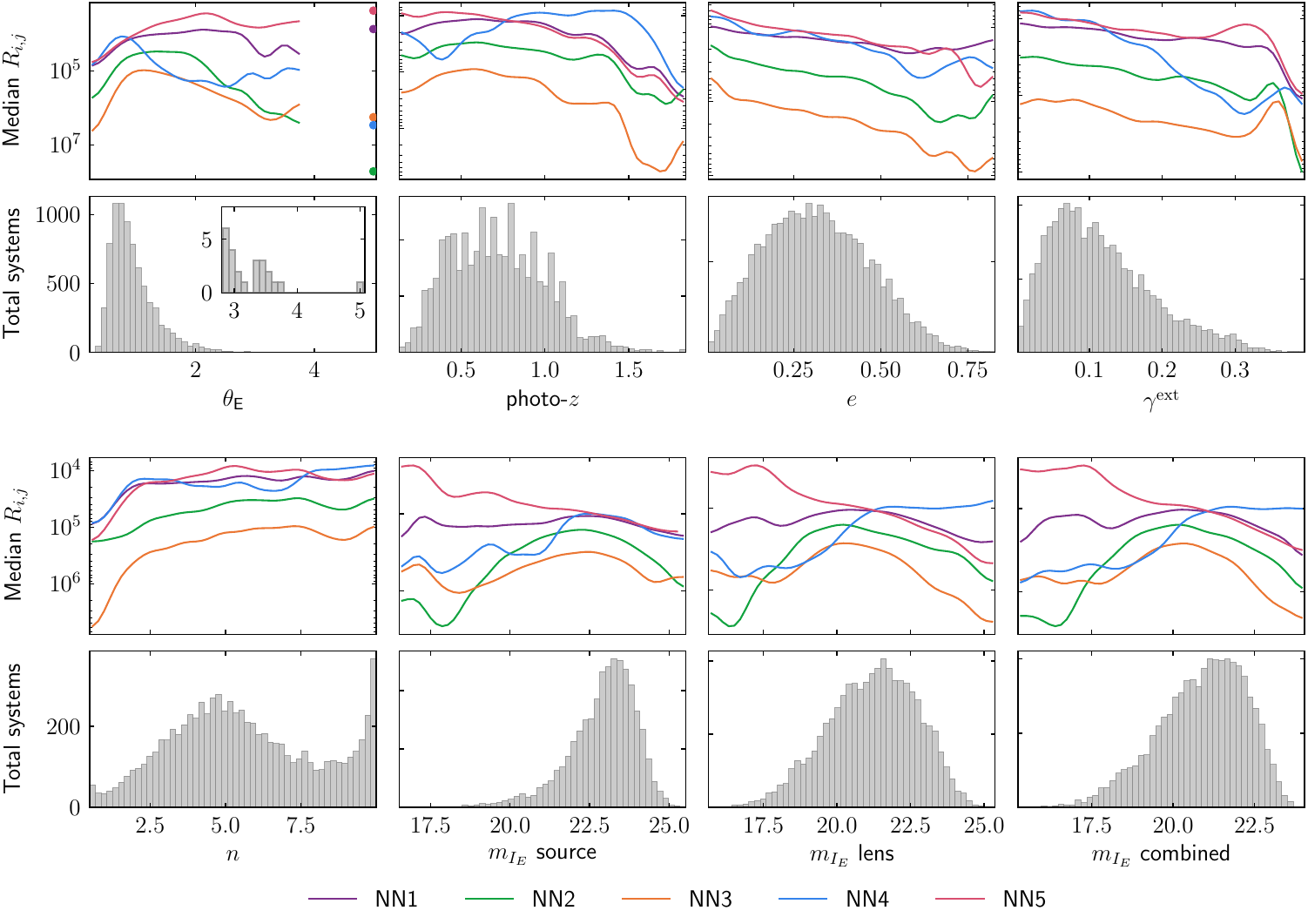}
\caption{Median global rank $R_{i,j}$ as a function of lens properties for the five networks. The upper row shows, from left to right, Einstein radius ($\theta_\sfont{E}$), photo-$z$, ellipticity ($e$), and external shear magnitude ($\gamma^{\rm ext}$). The bottom row shows the lens S\'ersic index ($n$) and the \IE-band magnitudes of the source, lens, and combined system. For each property, the upper panel shows the Gaussian-smoothed median $R_{i,j}$, with higher positions on the $y$-axis corresponding to higher rankings. The lower panel shows the number of systems in each of the 50 equal-width bins. The inset in the $\theta_\sfont{E}$ histogram provides an enlarged view of the sparsely populated region.}
\label{fig:bias_panels}
\end{figure*}

\subsection{\label{sc:NN1}NN1 -- \texttt{AstroVink}}
NN1 is a ViT based network named \texttt{AstroVink} \citep{Q1-Vincken}. Specifically the \texttt{AstroVink-DR1} variant is used. This network is pre-trained on natural images using the DINOv3 self-supervised framework \citep{Simeoni2025} and fine-tuned in stages on strong lensing data to manage the domain shift between simulations and real \Euclid observations. Input augmentations including corner cropping, rotations, and flipping are kept deliberately mild to avoid distorting lensing features. However, the various orientations of input data in combination with the ViT architecture allow detections of off-centered and complex configurations. Early stopping is applied using validation AUC, prioritising ranking performance over classification error. This network takes \IE-only JPEG cutouts as input.

\subsection{\label{sc:NN2}NN2 -- \texttt{OU-100-R8}}
NN2 is an ensemble of eight CNNs sharing the same architecture, each trained on a differently augmented version of the data using rotations of $90\,\mathrm{deg}$ and flipping. Predictions are averaged across the ensemble, with the standard deviation providing a per-object uncertainty estimate; this approach is found to reduce false positives from artefacts and stars. Weights are initialised from a previous search \citep{Wilde_2022} and fine-tuned on the common data pool. Early stopping is applied using validation loss, prioritising overall prediction accuracy over ranking performance. This network takes \IE-only JPEG cutouts as input.

\subsection{\label{sc:NN3}NN3 -- \texttt{4-layer CNN}}
NN3 is a four-layer CNN \citep{Pearce-Casey24}, the only network with no pre-trained initialisation. Training consists of two stages: the first uses the simulation sets, with class weights applied to handle the class imbalance; the second stage refines the network on real \Euclid candidates. No early stopping was applied and the final epoch weights are used. This network also takes \IE-only JPEG cutouts as input.

\subsection{NN4 -- \texttt{Zoobot}}
NN4 uses the \texttt{Zoobot} framework \citep{walmsley2024scalinglawsgalaxyimages, Q1-SP053} with the ConvNeXT-Tiny encoder pre-trained on morphological labels from the GalaxyZoo (GZ) project \citep{2025arXiv251223691W}. Unlike the other networks, it is trained as a regressive network predicting a continuous lens score rather than a binary classifier, with training examples re-weighted using previously collected inference scores. This means the grade~A samples receive a higher weight then the grade~B samples. Early stopping is applied using validation loss, prioritising overall prediction accuracy over ranking performance. This network also takes \IE-only JPEG cutouts as input. 

\subsection{NN5}
NN5 also uses the \texttt{Zoobot} framework, specifically the ConvNeXT-Base encoder trained on the morphological labels from GZ. It is an ensemble of two models, one trained on \IE-only data and the other on \IE+\JE+\HE, giving it access to near-infrared (NIR) colour information. This network uses its data in FITS format, as JPEG encoding is found to degrade performance when combining bands. Early stopping is applied using validation PR-AUC, prioritising ranking performance over classification error.

\section{\label{sc:ML-behaviour}Network behaviour results}
In this section we analyse how the networks behave when ranking our 50~deg$^2$ DR1 calibration set. Since all five networks have been applied to this calibration set, we can compare their behaviour on equal footing before applying them to the full sample. The lens candidates in our calibration set consist of a total of 627 grade~A and grade~B high-confidence systems.

\subsection{\label{sc:ranking-results}Ranking consensus and recovery}
Figure~\ref{fig:ranking-agreement} shows how the five networks compare in their ranking of candidates. We analyse the recovered grade~A and grade~B systems within each network's top 10\,000 candidates and the Spearman rank correlation coefficients between all pairs of networks.

The overlap matrix shows that NN5 recovers the most grade~A and grade~B systems (422), followed by NN1 (399), NN4 (272), NN2 (217), and NN3 (153). NN1 and NN5 share the largest number of detections (317), while NN3 and NN4 share the fewest (98). Despite NN1 and NN5 individually recovering the most systems, a substantial fraction of their detections are not found by the other networks, highlighting that high individual recovery does not guarantee overlap.

The Spearman correlation matrix reveals that no pair of networks is strongly correlated. The highest correlation is between NN2 and NN3 ($r_\sfont{S} = 0.61$), followed by NN1 and NN5 ($r_\sfont{S} = 0.56$). The lowest correlation is between NN1 and NN2 ($r_\sfont{S} = 0.21$), indicating that these two networks rank objects in very different ways. All correlations are positive, meaning no pair of networks fully disagrees, but the generally low values ($r_\sfont{S} < 0.5$ for most pairs) confirm that the networks provide complementary rankings. Notably, NN2 and NN3 show the highest ranking correlation ($r_\sfont{S} = 0.61$) but only share 110 systems in their top \num{10000}. This shows that similar ranking behaviour does not always translate into identical object selections, with each network favouring its own subset of systems despite broad agreement in how they order candidates.

Besides analysing ranking behaviour, we briefly look at the raw score distributions per grade in Appendix~\ref{fig:score_distributions}. These distributions show how each network assigns confidence to different grades, but they do not directly reflect candidate selection in practice, since VI is performed on the top-ranked objects regardless of absolute scores. The distributions show large differences between networks. NN1 assigns many grade~A systems scores close to 1.0 but a notable amount of grade~X also have high scores. NN2 assigns high scores ($\sim1.0$) rarely, but these systems have high purity. NN3 is more spread out across all grades. NN4 has the highest purity when looking at scores near 1.0, with grade~A completely dominating. NN5 also has many grade~A systems near 1.0 but they are accompanied by grade~X at high scores. Grade~B and grade~C follow the shape of grade~X in NN2 and NN3. In NN4 they decline from low to high scores. In NN1 and NN5 they are spread across the range with slightly higher numbers near 1.0.

\subsection{\label{sc:ensemble-results}Union recovery}
Figure~\ref{fig:combination_recovery} shows the number of lenses recovered as a function of the number of networks combined. We evaluate this at three different selection depths, $N = 1000$, $2000$, and $\num{10000}$, to analyse performance at different inspection budgets. To visualize the distribution of recovered systems we use box plots.

At all three selection depths, adding more networks to the union increases the total number of recovered systems. At $N = 1000$, a single network recovers between 121 and 337 systems depending on the network, while the union of all five recovers 492. At $N = 2000$, single-network recovery ranges from 153 to 422, increasing to 569 for the full five-network union. At $N = \num{10000}$, single-network recovery ranges from 276 to 534, and the five-network union recovers 623, approaching the total number of lenses in the calibration set. The largest gains come from adding a second or third network. At $N = 1000$, the median recovery increases from 209 for a single network, to 368 for a pair, and 413 for three networks. The spread in recovery narrows as more networks are combined, as each additional network covers systems missed by the others.

Figure~\ref{fig:leave_one_out} shows the leave-one-out analysis at $N = 2000$. NN5 makes the largest unique contribution ($\Delta_i = 51$), followed by NN1 ($\Delta_i = 46$) and NN4 ($\Delta_i = 30$). NN2 and NN3 contribute fewer unique systems, with $\Delta_i = 18$ and $\Delta_i = 8$, respectively. When NN3 is excluded, the remaining four networks still recover 561 of the 569 systems found by all five, while excluding NN1 drops recovery to 523. This shows that while all networks add value, NN1, NN4, and NN5 are the most complementary to the rest of the ensemble.

\subsection{\label{sc:performance-results}Performance metrics}
Figure~\ref{fig:roc_pr_combined} shows the ROC and precision-recall curves for the networks on the calibration set. NN5 achieves the highest AUC (0.99), followed by NN4 (0.97), NN1 (0.96), NN2 (0.92), and NN3 (0.88). The union of all five networks, computed by averaging their scores, also reaches an AUC of 0.99.

The precision-recall curves reveal larger differences between the networks. NN4 stands out with the highest precision at low recall, maintaining near-perfect purity in its top-ranked candidates. NN1 and NN5 also maintain high precision at low recall, while NN2 and NN3 drop more quickly. At higher recall, NN5 and NN1 recover more systems before precision begins to decline. In the highest purity range, the union curve drops below the best individual network (NN4), indicating that for small inspection budgets this network alone will yield the most true lenses. At larger inspection budgets the union outperforms individual networks, as slightly lower purity can be tolerated in exchange for higher completeness.

\section{\label{sc:Characterisation}Characterisation of detections}
In this section we present the results of applying the networks to the DR1 inference set. Following the selection function, ML classification, citizen science, and expert grading, the final catalogue contains 26\,662 classified objects. Of these, 6\,073 are classified as grade~A, 8\,999 as grade~B, and 9\,010 as grade~C, giving a total of 24\,082 strong lens candidates. The remaining 2\,580 objects are classified as grade~X (non-lenses). The combined grade~A and B sample we use for the analysis in this section contains a total of 15\,072 systems.

To study the biases the NN have towards physical properties of the strong lens candidates, we use the lens models produced by the \texttt{PyAutoLens} pipeline presented by \citet{DR1-SL03}. The model fits were visually inspected in a dedicated validation campaign, which also included the validation of segmentation maps and related model inputs presented by \citet{DR1-SL07}. The model panels were inspected by experts with experience in strong lens mass modelling. Each system required at least four votes from different experts, with at least 75\% agreement needed to classify a model as successful. This threshold is used consistently across the studies related to the ECLIPSE catalogue. In total, 4\,086 grade~A (67.28\%) and 3\,877 grade~B (43.08\%) systems were successfully modelled, giving a total of 7\,963 systems. Details on the modelling process, validation campaign, successful models, and failure modes are further described by \citet{DR1-SL03}.

\subsection{\label{sc:results-biases}Network biases over lens properties}
Figure~\ref{fig:bias_panels} shows the median global rank $R_{i,j}$ of the 7\,963 successfully modelled grade~A and grade~B systems as a function of eight lens properties: Einstein radius $\theta_\sfont{E}$, photo-$z$ of the lens, ellipticity ($e = 1-q$) where $q$ is the minor-to-major axis ratio, external shear magnitude $\gamma^{\rm ext}$, S\'ersic index $n$, and the \IE magnitudes of the source, lens, and combined system. For each property, we divide the systems into 50 equal-width bins and calculate the median $R_{i,j}$, with Gaussian smoothing applied to show the general trends. The lower panels show the number of systems per bin, since some bins contain very few or no systems.

The Einstein radius $\theta_\sfont{E}$ shows that NN1 and NN5 assign the highest rankings to large systems. The other networks show a stronger decline beyond $\theta_\sfont{E} \sim 1\farcs5$ to $2\arcsec$. Most systems have $\theta_\sfont{E} < 1\farcs5$, and only a small fraction extends beyond $3\arcsec$. These larger systems could potentially be group-scale rather than galaxy-scale lenses. Since only the highest-ranked candidates are selected for visual inspection, the lower rankings assigned by NN2, NN3, and NN4 could cause some of these larger systems to be missed.

The photo-$z$ estimates come from the pipeline described by \citet{DR1-SL04}. Reliable estimates of the deflector galaxy are available for 2\,555 grade~A (62.53\%) and 2\,250 grade~B (58.03\%) systems with successful models, leaving 4\,805 systems for this panel. NN4 shows a slight preference for systems around $z \sim 0.8$ to $1.4$, assigning them higher rankings than systems at lower photo-$z$. This preference is not seen as clearly in the other networks, suggesting that NN4 could help recover higher-redshift lenses that might otherwise fall outside the candidate selection. NN3 shows a different preference, assigning its highest rankings around $z \sim 0.5$ to $0.6$ and lower rankings towards higher photo-$z$. NN1 and NN5 follow almost identical trends, suggesting that they have very similar selection biases for this property.

Most networks assign higher rankings to systems with lower ellipticity $e$ and external shear $\gamma^{\rm ext}$. Specifically for $e$, there does not seem to be a clear preference that stands out for any single network. For external shear, NN4 shows a strong decrease in rankings between $\gamma^{\rm ext} \sim 0.15$ and $0.3$, while NN5 shows a small increase in rankings around $\gamma^{\rm ext} \sim 0.3$. These differences may help NN1 and NN5 recover some systems with higher ellipticity or shear that the other networks rank lower, although the differences in the rankings are small.

The S\'ersic index $n$ shows that all networks assign lower rankings to systems with $n \lesssim 2$, with higher rankings towards higher $n$ values. Although the populations here are much smaller, there are still enough that we should not ignore the patterns completely. A slight preference is seen for NN2 and NN3, while the other networks show smaller changes beyond $n \sim 3$. This can suggests a shared bias towards more centrally concentrated lens galaxies, meaning that systems with low $n$ may be missed even when combining all the networks.

The \IE magnitude panels show different preferences between the networks. NN5 clearly assigns higher rankings to bright systems (in regards of lens, source and combined system), while NN4 shows a preference for fainter systems. NN2 and NN3 show a different pattern, assigning higher rankings to systems with intermediate lens and combined magnitudes around $m_{\rm \IE} \sim 20$ to $22$, with lower rankings at both ends of the range. NN1 shows smaller changes across most of the magnitude distributions, without a clear preference for bright or faint systems. These differences suggest that combining the networks could help recover lenses across a wider range of magnitudes.

\subsection{\label{sc:results-unique}Unique detections and false positives per network}
\begin{figure*}
    \centering
    \includegraphics[width=\textwidth]{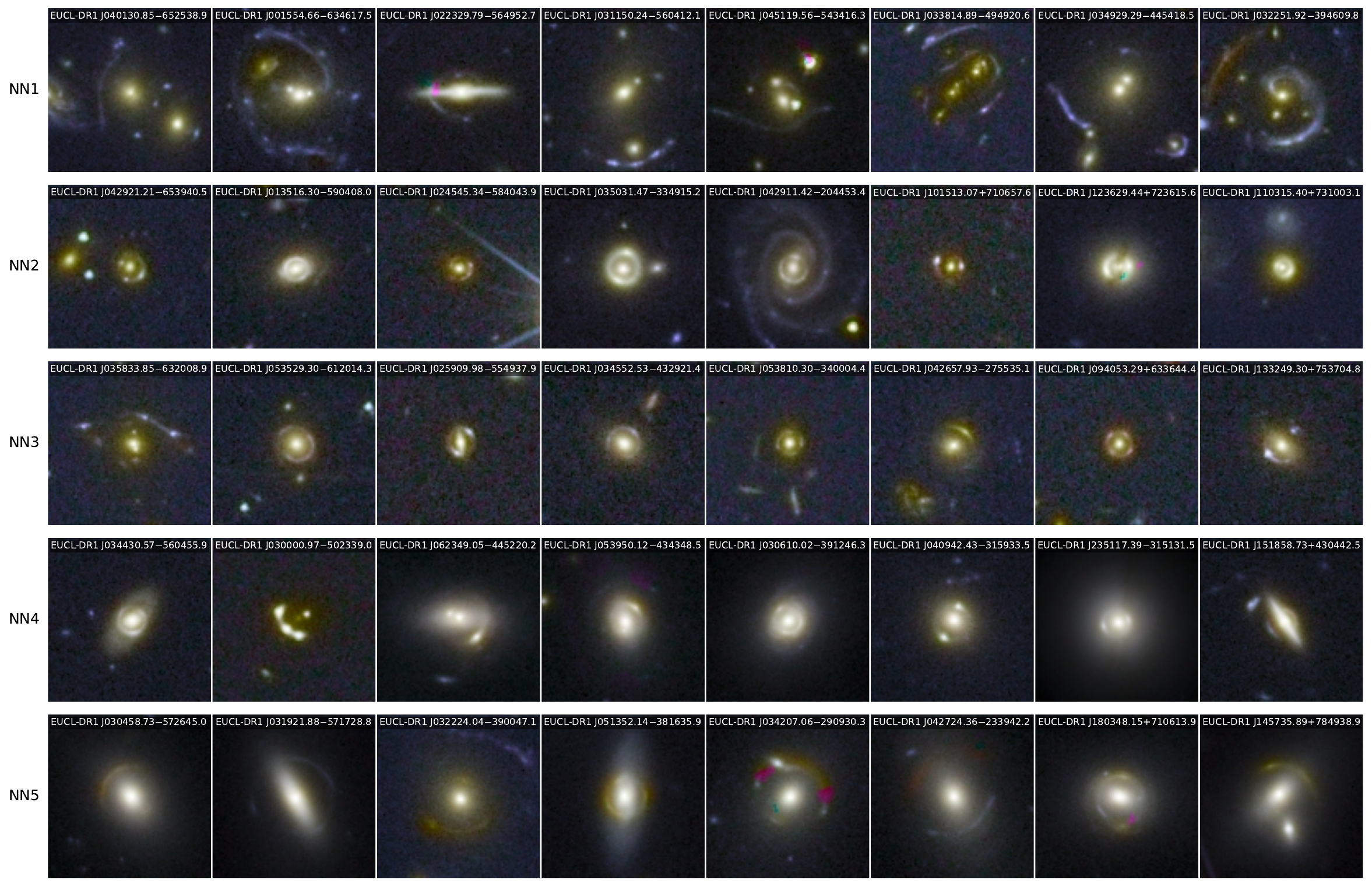}
    \caption{Mosaic with $10\arcsec$ cutouts of unique grade~A lens detections per network from the DR1 catalogue. Each row shows examples of grade~A systems that were uniquely recovered by a single network. These systems were not found in the top 2000 highest ranks of any other network.}
    \label{fig:unique_mosaic}
\end{figure*}

\begin{figure*}
    \centering
    \includegraphics[width=\linewidth]{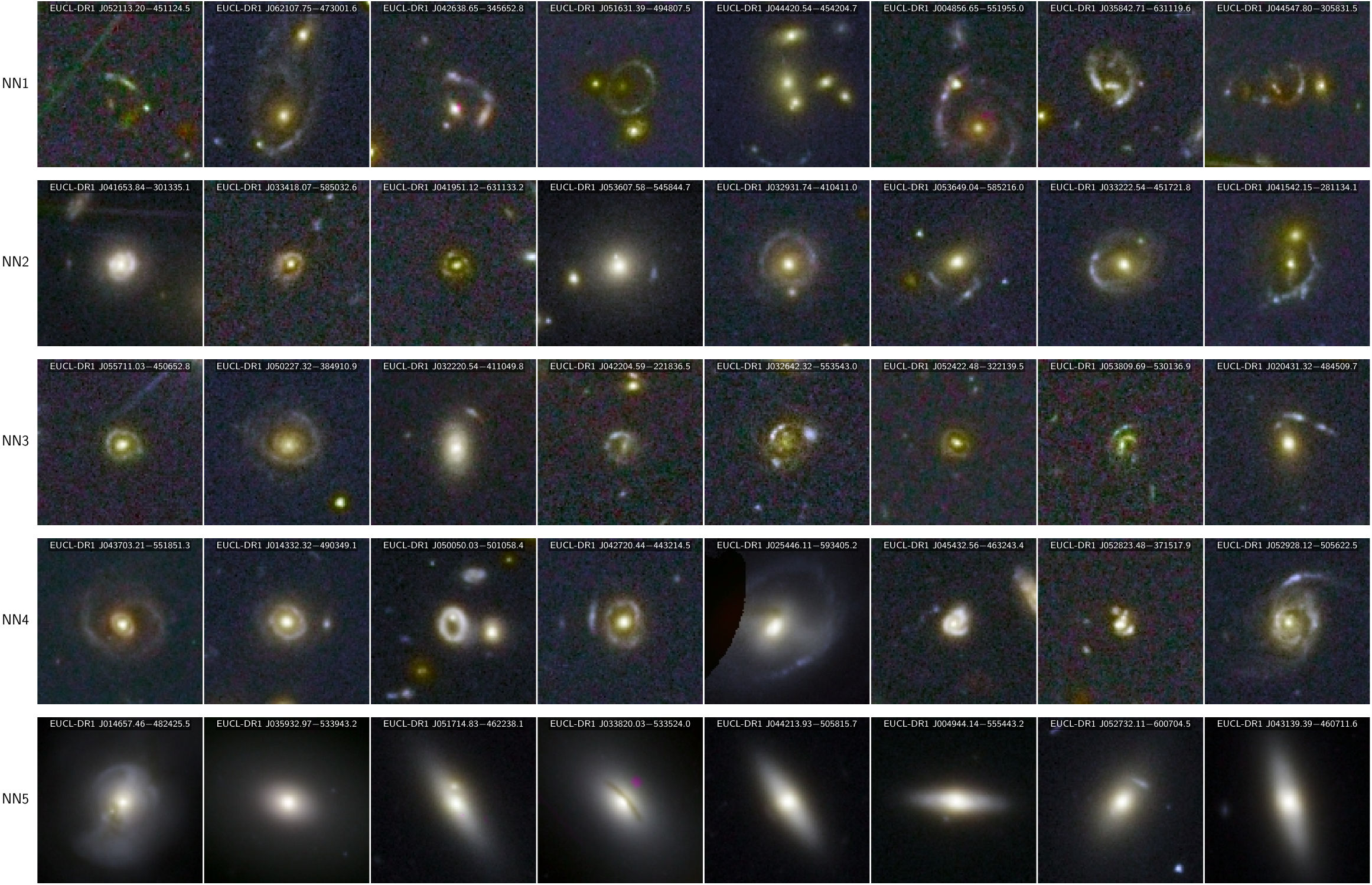}
    \caption{Mosaic with $10\arcsec$ cutouts of unique false positives per network from the DR1 catalogue. Each row shows examples of grade~X systems that were uniquely recovered by a single network. These systems were not found in the top 2000 of any other network.}
    \label{fig:false_positives}
\end{figure*}

Figure~\ref{fig:unique_mosaic} shows examples of grade~A candidates found within the top 2000 candidates of one network, but not within the top 2000 of any other network. Additionally, their unique grade~X (false-positives) are shown in a similar way in Fig.~\ref{fig:false_positives}. This lets us visualize the differences in the types of systems found by each network.

\begin{figure}
    \centering
    \includegraphics[width=1\linewidth]{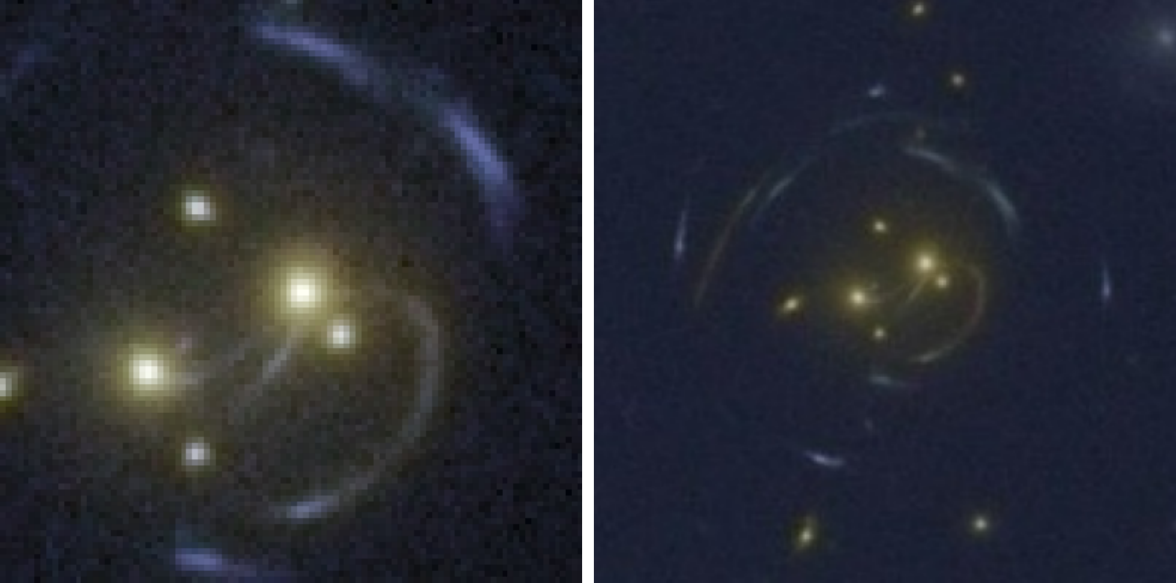}
    \caption{Strong lens candidate EUCL-DR1\,J024809.51$-$395548, identified by NN1 and NN5 in their top 2000 highest ranks. The left panel shows the original $10\arcsec$ cutout, while the right panel shows a larger $20\arcsec$ cutout needed to capture the full system.}
    \label{fig:nn1_fifth_system_side_by_side}
\end{figure}

Figure~\ref{fig:bias_panels} already indicated that NN1 may have a preference towards larger $\theta_\sfont{E}$ systems. This is also clearly shown in some of the unique lens candidates it found, where we can see several large systems with extended arcs and complex configurations. Some of the systems found by this network are likely to be group- or cluster scale. One of them, EUCL-DR1\,J024809.51$-$395548, is highlighted in Fig.~\ref{fig:nn1_fifth_system_side_by_side}. Here a larger cutout of $20\arcsec$ shows the extension of the system beyond what was originally detected in the $10\arcsec$ cutouts the network was applied to. When looking at the grade~X systems found by NN1 we also see some slightly larger extended features or multiple nearby galaxies, which may explain why the network selected them.

NN2 shows a different pattern, with several compact systems containing small arcs close to the deflector. Some of its grade~X detections show similar features, including ring galaxies and spirals near the center. NN3 also finds several compact lenses, although the arcs are often smaller or less visible. Its false positives include similar objects with partial rings or spiral features. From these examples, NN2 and NN3 appear to find similar types of systems, without a clear difference that stands out between them.

NN4 finds several systems with very clear lensing features. Some examples also show multiple arcs that could indicate double-source plane lensing. This may be a result to the training approach of NN4, where grade~A systems received higher weights than grade~B systems, possibly favouring systems with more obvious lensing features. The grade~X detections also include several ring and spiral galaxies with similar structures.

\begin{figure}
    \centering
    \includegraphics[width=1\linewidth]{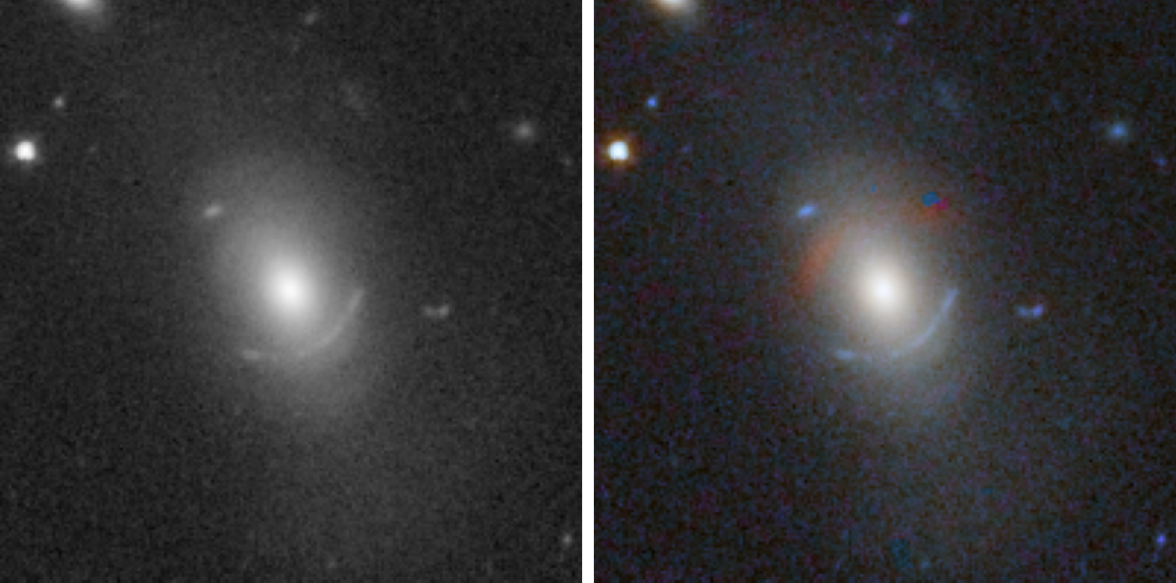}
    \caption{Strong lens candidate EUCL-DR1\,J042724.36$-$233942.2, identified by NN5. The left panel shows the \IE band only, while the right panel combines the \IE band with the \JE and \HE bands of the near-infrared spectrum, revealing a potential second lensed source that is not immediately visible in the \IE-only image.}
    \label{fig:red_side_by_side}
\end{figure}

Finally, NN5 finds several systems with bright central galaxies and arcs that appear more red. This may be related to NN5 being the only network using the NIR bands, allowing it to identify lensing features with different colours. One example of this is EUCL-DR1\,J042724.36$-$233942.2, as shown in Fig.~\ref{fig:red_side_by_side}. Here combining the \IE band with the \JE and \HE bands reveals a potential second lensed source that is not immediately visible in the \IE-only image. The bright central galaxies are also consistent with the preference for lower lens magnitudes found in Fig.~\ref{fig:bias_panels}. The grade~X detections for NN5 stand out compared to the other networks, with many smooth elliptical and bright galaxies. This may suggest that NN5 also uses information from bright foreground galaxies when identifying candidates.

\section{\label{sc:DiscussionConclusion}Discussion}
As the title of this paper suggests, learning about machine learning is as important as applying it. Understanding why networks find what they find, and what they miss, is essential for building complete catalogues from automated searches. This work provides an approach for doing so by combining ranking correlations, union recovery analysis, and property-dependent bias characterisation, we can evaluate not just how well a set of classifiers performs, but where they fail and how to correct for it.

We applied all networks to a 50~deg$^2$ calibration set and later to a selection of approximately 68 million objects. Our analysis shows that the networks are complementary, not redundant. Despite having access to the same training data, each network develops its own biases driven by differences in architecture, image processing, and training strategy. In the calibration set, NN5 recovers the most grade~A and grade~B systems, followed by NN1, while NN4 achieves the highest precision at low recall. The leave-one-out analysis further shows that NN1, NN4, and NN5 contribute the largest number of unique systems.

To understand what types of systems the networks recover and how this affects the completeness of our catalogue, we analysed their lens properties and unique detections across DR1. One interesting finding is that NN1 and NN5 both favour larger Einstein radii. NN1 uses a ViT with corner-cropping augmentation, which may explain its ability to detect extended arcs and off-centred systems. Its unique detections also show multiple complex configurations, including systems extending beyond the original $10\arcsec$ cutouts. NN5 also ranks larger systems highly, but its unique detections include several systems with bright central galaxies and redder arcs, while NN1 stands out for its extended and complex configurations. NN2 and NN3 show a different pattern, favouring smaller arcs close to the deflector. Their similar ranking behaviour is reflected in their relatively high correlation ($r_\sfont{S}=0.61$) and smaller unique contributions.

The image processing and training strategies may help further explain the differences between networks. For example, NN4 gives higher weights to grade~A systems during training, which may contribute to its high precision and preference for systems with clear lensing features. NN5 uses CLAHE contrast enhancement and Sobel edge detection, which could make foreground galaxy structures more prominent. This processing could potentially explain its preference for brighter lens candidates and the bright elliptical galaxies in its false-positives, although this would need further testing to confirm. Explainable AI methods like Grad-CAM \citep{Selvaraju_2019} can also help determine whether the network focuses mainly on the foreground galaxy or the lensing features. NN5 also uses combined \IE+\JE+\HE images, which may help identify sources with different colours. A more detailed analysis on the impact of image processing and training data selection is presented by \cite{DR1-SL06}.

Some biases are shared across all networks. One example is their preference for higher S\'ersic indices, which could cause disk-dominated lenses to be missed. This may come from the training data, where many simulated lenses use luminous red galaxies as deflectors. Combining more networks with this same preference would not necessarily improve completeness. Instead, future training could include more examples of disk-dominated systems, such as those investigated by \Ecker{}. 

Another limitation is that our analysis depends on both the ML selection and the subsequent VI. Lens populations that are poorly represented in the training data may be ranked too low to reach VI, while true lenses that do reach VI may still be missed or misclassified. Both effects can potentially also reduce the completeness of the recovered lens population. Furthermore, the lens-property analysis only includes successfully modelled systems, which may not fully represent the properties of the entire catalogue. The current search strategy selects the top-ranked candidates from each network without considering what population those candidates represent. Since less than $0.1\%$ per network out of the ranked objects is passed on to VI, this can cause certain lens types to be under-represented in the final catalogue. For upcoming \Euclid releases, the biases identified in this work could be used to select different numbers of candidates from each network based on the lens populations they recover. For example, NN1 could be given larger cutouts for inference, while NN5 could provide systems with a wider range of colours. NN4 could provide a high-purity selection, particularly when the VI budget is limited. By combining these selections and accounting for their overlap, we could use the available VI budget to recover a wider range of lens types, rather than only maximising the total number of detections.

The ECLIPSE catalogue provides thousands of lens candidates across a wide range of properties, which is an order of magnitude more data than was available for the current generation of networks. As we have seen in previous work \citep{Lines_2025, Q1-Vincken}, retraining networks on this expanded sample will improve the completeness of the recovered population, as the networks will have seen a broader variety of real lens configurations during training. Retraining therefore may increase the completeness of the catalogue in later stages, especially if the retraining focuses on the strengths and biases we learned about the networks. The unique false positives could also be useful for future training, particularly the ring and spiral galaxies selected by NN2, NN3, and NN4, and the bright elliptical galaxies selected by NN5. These could be added as targeted negative examples to reduce similar false positives in future searches.

Finally, the analysis described in this work is not only useful to \Euclid. Upcoming wide-field surveys such as LSST and \textit{Roman} will face the same challenge of selecting rare objects from billions of sources using multiple classifiers. The approach we present here can serve as a standard for how to evaluate and combine classifiers in any large-scale search, ensuring that the resulting catalogues are not just large but also representative of the true underlying population.

\begin{acknowledgements}
\AckEC  
\AckDRone
FC is supported by the Agencia Estatal de Investigaci\'on (AEI), Ministerio de Ciencia, Innovaci\'on y Universidades, Spain, under the project PID2024-155455NB-I00, within the 2024 Call for Knowledge Generation Project and by project CNS2023-144016, and co-funded by the European Union -- NextGenerationEU.
\end{acknowledgements}
%
%

\bibliography{Euclid, Q2, my} 
%
%

%

\begin{appendix}
\onecolumn
\begin{multicols}{2}
\section{Score distributions}
This appendix presents the detailed score distributions for each network across all grade categories, providing insight into how confidently each network separates lens candidates from contaminants.
\nolinenumbers
\\\\\\\\\\
\end{multicols}
\begin{figure}
    \centering
    \includegraphics[width=\columnwidth]{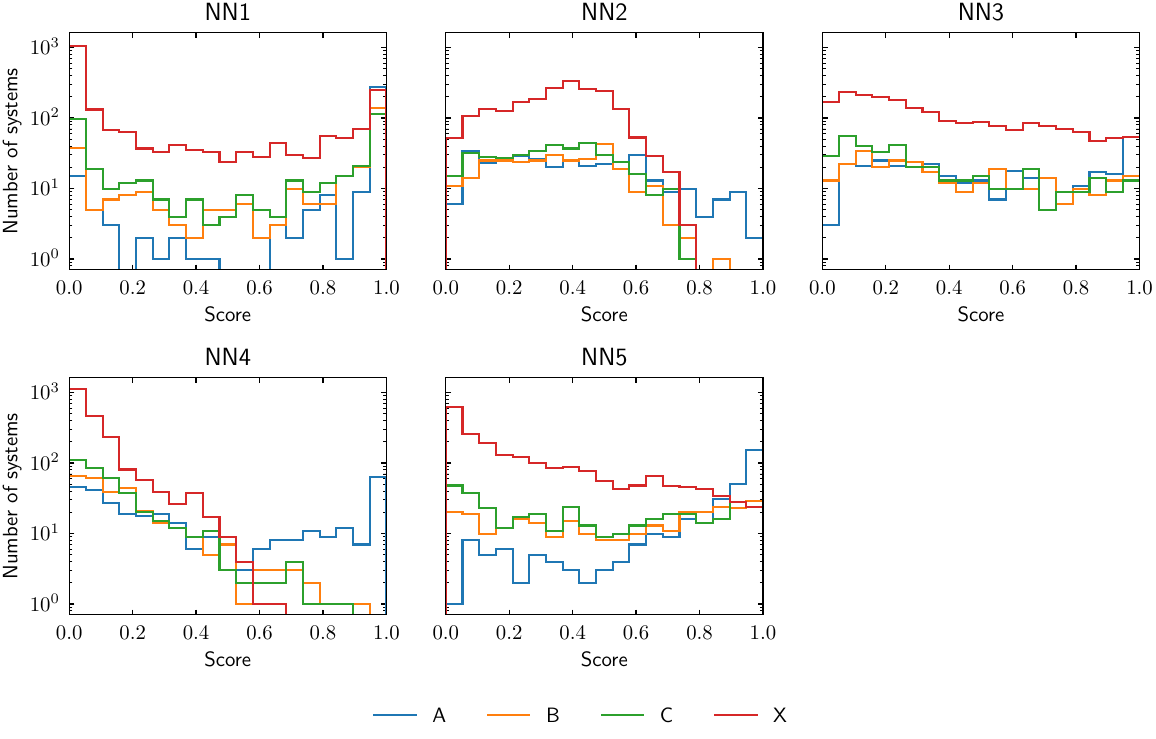}
    \caption{Score distributions for each network, split by grade (A, B, C, X). Each panel shows a histogram of predicted scores on the $x$-axis ranging from 0 to 1 against the number of objects on the $y$-axis (logarithmic scale). The colours indicate the different grades, with blue for grade~A, orange for grade~B, green for grade~C, and red for grade~X.}
    \label{fig:score_distributions}
\end{figure}
\end{appendix}

\end{document}